\documentclass{aa} 

\usepackage{txfonts}
\usepackage{natbib}
\usepackage{soul}
\usepackage{xcolor}
\usepackage{multicol}
\usepackage{graphicx,graphics,rotating,amssymb,amsmath}
\usepackage{xcolor}
\usepackage{multirow}
\usepackage{longtable}
\usepackage{booktabs}
\usepackage{subfig}
\usepackage{float,capt-of}
\usepackage{natbib}
\usepackage[english]{babel}
\usepackage{morefloats}
\usepackage{color}
\usepackage{booktabs}
\usepackage{sidecap}
\usepackage{epsfig,color}
\usepackage{wrapfig}
\usepackage{lscape}
\usepackage{ulem}
\usepackage{url}
\usepackage{tablefootnote}
\usepackage{hyperref}
\newcommand{\herschel}{{\it Herschel}}
\newcommand{\nthp}{N$_2$H$^+$ (1-0)}

\usepackage{graphicx}
\usepackage{stfloats} 
\begin{document}

   \title{Emergence of high-mass stars in complex fiber networks (EMERGE)}

   \subtitle{II. The need for data combination in ALMA observations}

   \author{Francesca Bonanomi
          \inst{1}
          \and
          Alvaro Hacar\inst{1}
          \and
          Andrea Socci\inst{1} 
           \and
          Dirk Petry\inst{2} 
          \and
          S{\"u}meyye Suri\inst{1} 
          }

    \institute{Institute for Astronomy (IfA), University of Vienna,
              T\"urkenschanzstrasse 17, A-1180 Vienna\\
              \email{francesca.bonanomi@univie.ac.at}
             \and
             European Southern Observatory, Karl-Schwarzschild-Strasse 2, 85748 Garching, Germany
             }
   \date{Received 12.12.2023; accepted 14.05.2024}

\titlerunning{EMERGE: II. The need for data combination in ALMA observations}


  \abstract
   {The Atacama Large Millimetre Array (ALMA)’s high-resolution images allow to resolve the filamentary structure of the Interstellar Medium (ISM) down to a resolution of few thousand au in star-forming regions located at kpc distances. 
   }
    {We aim to systematically quantify the impact of the interferometric response and the effects of the short-spacing information
    during the characterization of the ISM structure using ALMA observations.  
   }
    {We create a series of continuum ALMA synthetic observations to test the recovery of the fundamental observational properties of dense cores and filaments (i.e. intensity peak, radial profile, and width) at different spatial scales. We homogeneously compare the results obtained with and without different data combination techniques using different ALMA arrays and SD telescopes in both simulated data and real observations. 
   }
    {Our analysis illustrates the severity of interferometric filtering effects. ALMA-12m alone observations show significant scale-dependent flux losses that systematically corrupt ($>$~30\% error) all the physical properties inferred in cores and filaments (i.e. column density, mass, and size) well before the maximum recoverable scale of the interferometer. These effects are only partially mitigated by the addition of the ALMA ACA-7m array although at the expenses of degrading the telescope point-spread-function (PSF). Our results demonstrate that only the addition of the ALMA Total Power information allows to recover the true sky emission down to few times the ALMA beamsize with satisfactory accuracy ($<$~10\% error).
    Additional tests demonstrate that the emission recovery of cores and filaments at all scales is further improved if the 7m+TP data are replaced by additional maps obtained by a larger SD telescope (e.g., IRAM-30m), even if the latter are noisier than expected. The above observational biases particularly affect partially resolved targets, becoming critical especially for studies in nearby regions such as Taurus or Orion.
   }
    {Our results demonstrate the need for the use of the state-of-the-art data combination techniques to accurately characterize the complex physical structure of the ISM in the ALMA era. 
   }


   \maketitle
%
\section{Introduction}\label{sec:intro}

The advent of the Atacama Large Millimetre Array (ALMA) has revolutionized the study of the Interstellar Medium (ISM) with its unprecedented high sensitivity and resolution. A decade of \herschel~observations have highlighted the complex filamentary structure of molecular clouds at parsec scales \citep{2014Andre}. Within these filaments recent ALMA observations have unravelled an unexpected physical and kinetic complexity at sub-parsec scales  \citep[e.g.][]{2013Peretto,2018Hacar,Chen2019,Li2021,Sato2023,Cunningham2023} drawing the picture of a hierarchical ISM \citep{2024Hacar}.

Resolving the sub-pc filamentary structure of star-forming regions located at kpc distances can only be performed using high-resolution interferometric observations. However, interferometric observations alone are strongly affected by intrinsic spatial filtering effects \citep[e.g.][]{2016Ossenkopf-Okada}. Interferometry in the radio and sub-millimeter regimes relies on the aperture synthesis technique \citep{1958Jennison}. The limited observing time leads to a sampling of the Fourier (\textit{u,v}) plane which is always discrete, irregular and incomplete.  The resulting interferometric visibilities do not contain information at all spatial scales but only of those Fourier components sampled by the specific baselines (distance between two antennas) available during the observation. By construction, the spatial sensitivity achieved using the aperture synthesis technique is limited by the shortest distance (baselines) between antennas within the array, roughly corresponding to the antenna diameter in modern interferometers such as ALMA.
The lack of these short baselines is commonly known as short-spacing problem and critically affects the recovery of extended emission at large angular scales compared to the interferometric beam size \citep{1994Wilner,2009Kurono}.

The solution to the short-spacing problem is to add additional observations  sampling large-scales, usually provided by a single-dish (SD) telescope, to be combined with the interferometric data \citep{2002Stanimirovic}. 
This technique, known as data combination, allows to preserve the extremely high angular resolution obtained with the interferometer together with the large-scale sensitivity given by the SD data. Several methodologies and algorithms have been developed during the last decades to perform such data combination both in the Fourier and image planes \citep[see][for a recent summary of these techniques and their implementation]{2023Plunkett}. 


In interferometric observations, the large-scale sensitivity is evaluated through the Maximum Recoverable Scale (MRS). According to \citealt{1994Wilner}, the MRS is defined as the angular size for an input Gaussian visibility distribution at which the ratio between the observed source peak brightness in absence of the short-spacing data and its true peak brightness is $1/e$. In the case of ALMA observations the MRS is defined as the largest angular scale to which the instrument is still sensitive. The ALMA Technical Handbook provides as empirical formula MRS$\sim0.983\lambda/L_{min}$, where $\lambda$ is the observed wavelength and $L_{min}$ is the shortest baseline of the interferometer\footnote{The 5th percentile baseline length is used instead of the shortest baseline, to provide a more robust measure of these properties.} \citep{2023Cortes}. The MRS plays a crucial role in the observational setup: both the choice of the interferometric arrays to use and of the array configuration depend on the largest angular scale to observe and on its comparison with the MRS. Thus, a detailed analysis of this parameter is needed in order to investigate the impact of the short-spacing effect on the quality of the observations.

As shown in different Galactic surveys \citep[see e.g.][]{2010Andre,2010Molinari,2011Arzoumanian}, both filaments and cores are typically embedded in large amounts of cloud gas, showing a shallow emission profile with large contributions from the bright diffuse emission extending over several parsecs in size. 
The recovery of this extended emission determines key observational properties such as the radial profile, source size, and total mass of cores and filaments, and therefore becomes essential to interpret the physical characteristics of the ISM structure.
This cloud emission at large angular scales can easily exceed the MRS of most interferometers. 
The absence of large-scale information in interferometric-alone observations leads to an incomplete and misleading representation of the true sky emission and poses a critical challenge for both continuum and molecular line observations.

This work is part of the Emergence of high-mass stars in complex fiber networks (EMERGE) project \citep[see][hereafter Paper I]{2024Hacar}\footnote{EMERGE Project website: https://emerge.univie.ac.at/}. 
The EMERGE project will survey the internal gas organization in a series of high-mass star-forming regions using high-resolution ALMA observations combined with additional high-sensitivity SD data. In this paper, the second of its series (Paper II), we investigate the accuracy of the ALMA observations used to characterize our EMERGE survey (see also \citet{2024Socci}; hereafter Paper III). Beyond the combination techniques explored by \citet{2023Plunkett}, this work systematically quantifies the effect of the interferometric filtering on the recovery of different emission properties (flux, angular size, and radial dependence) used to derive the physical characteristics (mass, size, stability, evolutionary state,...) of relevant ISM structures such as cores and filaments. Since the lack of short-spacing information becomes apparent in resolved targets, the results of our work are most likely applicable to the study of nearby regions within 1~kpc such as Taurus and Orion typically observed using the most compact ALMA configurations (e.g., C43-1; see Paper I). Nonetheless, these results might be applicable to other Galactic targets if observed at high spatial resolution using more extended ALMA configurations.

We have evaluated the current use of data combination in ALMA observations in recent Cycles (Sect.~\ref{sec:archive}).
We produce and analyze synthetic ALMA observations (Sec.~\ref{sec:method}) to evaluate the ability of interferometric observations recovering fundamental emission properties of idealized cores (Sec.~\ref{sec:cores} and Appendix~\ref{sec:cmf}), and different filamentary geometries (Sect.~\ref{sec:fil}).  We test different data combination methods using the three ALMA arrays and evaluated the Point Spread Function (PSF) obtained form the observations (Sect.~\ref{sec:psf}).  We also investigate the combination of ALMA 12m only with a large SD such as the IRAM-30m telescope (Sect.~\ref{sec:iram}) and discuss the sensitivity requirements for data combination (Sect.~\ref{sec:noise}). Finally we compare the results obtained with and without data combination, and in between different data combination techniques, with real observations (Sect.~\ref{sec:real data}). 

\section{Data combination with ALMA}\label{sec:archive}

\begin{figure*}[htbp]
	\centering
	\includegraphics[width=1\columnwidth]{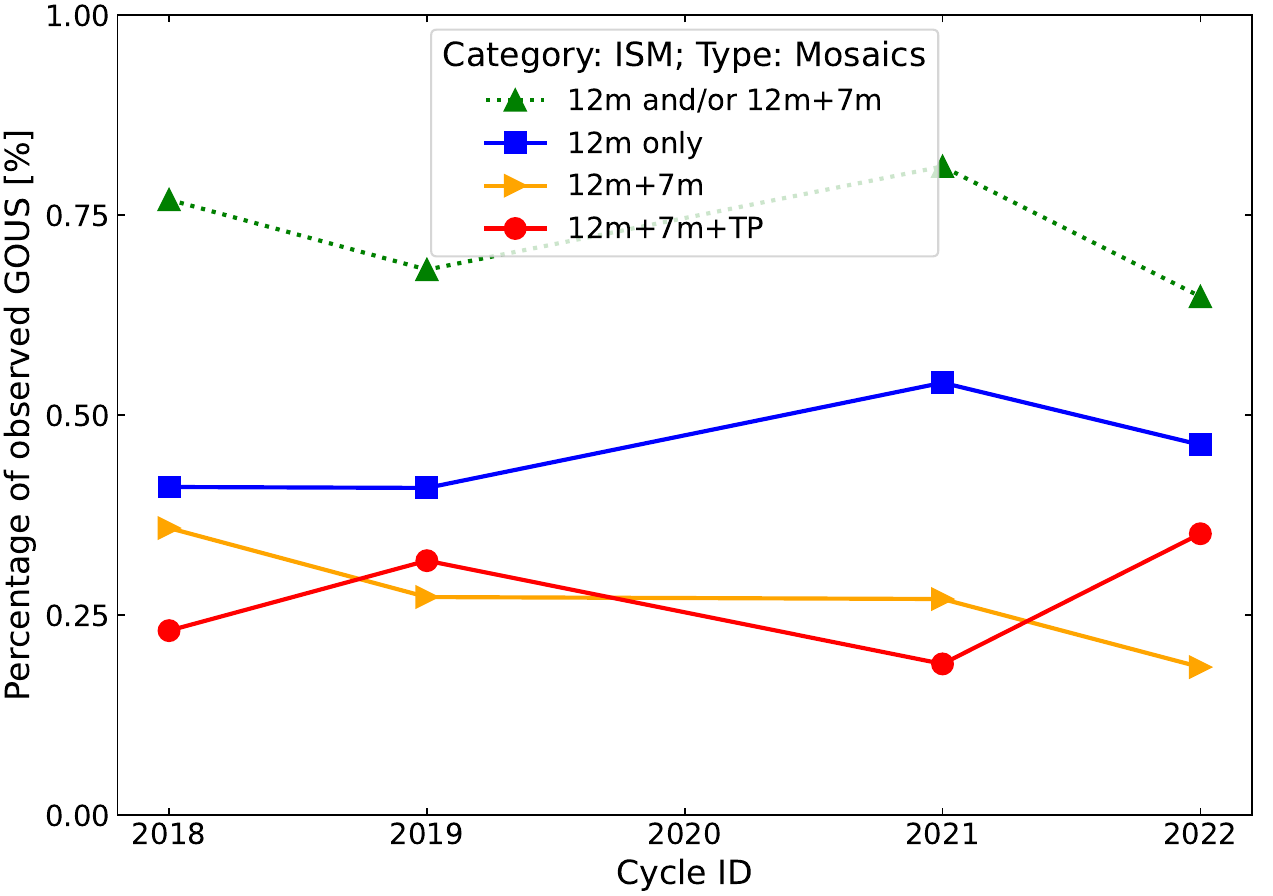}
        \includegraphics[width=1\columnwidth]{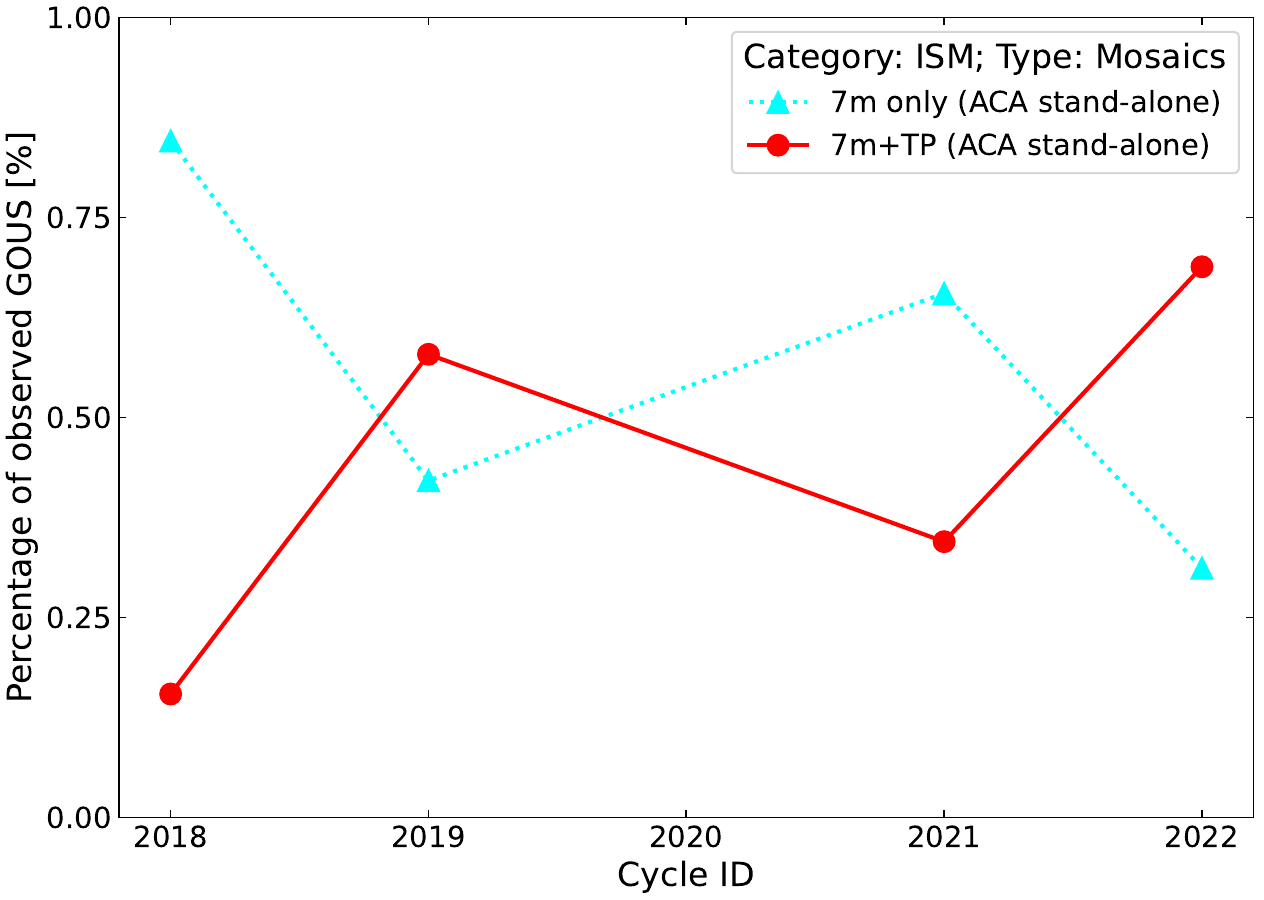}
	\caption{Observed GOUS with ALMA mosaics for the ISM category over ALMA Cycles 6-9 (years 2018-2022). 
    {\bf (Left panel)} Percentage of GOUS observed with ALMA obtained with the 12m (blue squares), 12m + 7m (orange triangles), and the 12m + 7m + TP (red circles) arrays, respectively. 
    The green triangles represents all the GOUS requested without SD observations (i.e. 12m and/or 12m+7m).
    {\bf (Right panel)} Percentage of GOUS observed with ACA in stand-alone mode including 7m-only (cyan triangles) and 7m+TP data (red circles).} Fluctuation of 10-20\% are expected from Poisson statistics given the typical number of GOUS considered per Cycle ($\sim$40 per year).
    
	\label{fig:alma-statistics} 
\end{figure*}

In order to improve its sensitivity at all angular scales, the ALMA observatory includes three arrays sampling different baseline distances: the ALMA 12m main array (composed by up to fifty 12m antennas arranged in distinct array configurations with baselines between 15 meters and 16 kilometers and sensitive to small angular scales), the ALMA Compact array (ACA)-7m array (twelve 7m antennas with baselines between 9 and 45 meters,  sampling the intermediate scales) and the Total Power array (TP, four SD 12m telescopes). Alternatively, the ALMA interferometric observations (12m alone or 12m + 7m) can be combined with a different SD, always considering the u-v coverage of the different arrays should ideally overlap for a reliable data combination.

Although the short spacing problem and the necessity of data combination have been well known in the community, even today the number of ISM projects aiming for combined 12m + 7m + TP array observations, and 7m + TP data in the case of ACA stand-alone observations, is lower than expected. To illustrate this issue, we queried the ALMA Science Archive using the ALminer toolkit \citep{2023Ahmadi}. We searched for individual Group Observing Unit Sets (GOUS) in ALMA Cycles 6-9 (years 2018-2022) and selected those completed projects within the ISM category requesting mosaic observations. Our conservative selection of mosaics ensures that these observations are aiming for targets with extended emission larger than the ALMA 12m (or 7m) primary beam, and therefore beyond the interferometric MRS, that most likely require data combination. 

We show the percentage of GOUS with 12m, 7m, and TP data per year in Fig.~\ref{fig:alma-statistics}. 
Even in the case of large 12m mosaics obtained with the ALMA main array (left panel), the number of interferometric alone observations (12m and/or 7m data; green triangles) accounts for $>$~60\% of the observed GOUS over these four Cycles, including $>$~40\% of GOUS requesting only 12m observations (blue squares) and $\sim$25-30\% of GOUS including 12m+7m data (orange triangles). In contrast, the number of GOUS requesting 12m+7m+TP (red circles) in the last years  has been consistently below $\sim50\%$ indicating that the severity of these effects has not been completely internalized. 
The situation is slightly better in the case of ACA stand-alone observations (right panel) showing an apparent increasing trend of ISM projects using ACA mosaics requesting 7m+TP GOUS in the last years (red circles), although $\sim$30\% of them still requested 7m-only data by 2022 (cyan triangles).
The results shown from this sub-sample can potentially affected a much larger number of ALMA projects using single 12m and 7m pointing observations in crowded galactic environments. Accordingly, the aim of this study is to quantify these effects to raise awareness in the community.

\section{ALMA simulations}\label{sec:method}

\begin{figure}[ht!]
	\centering
	\includegraphics[width=1\columnwidth]{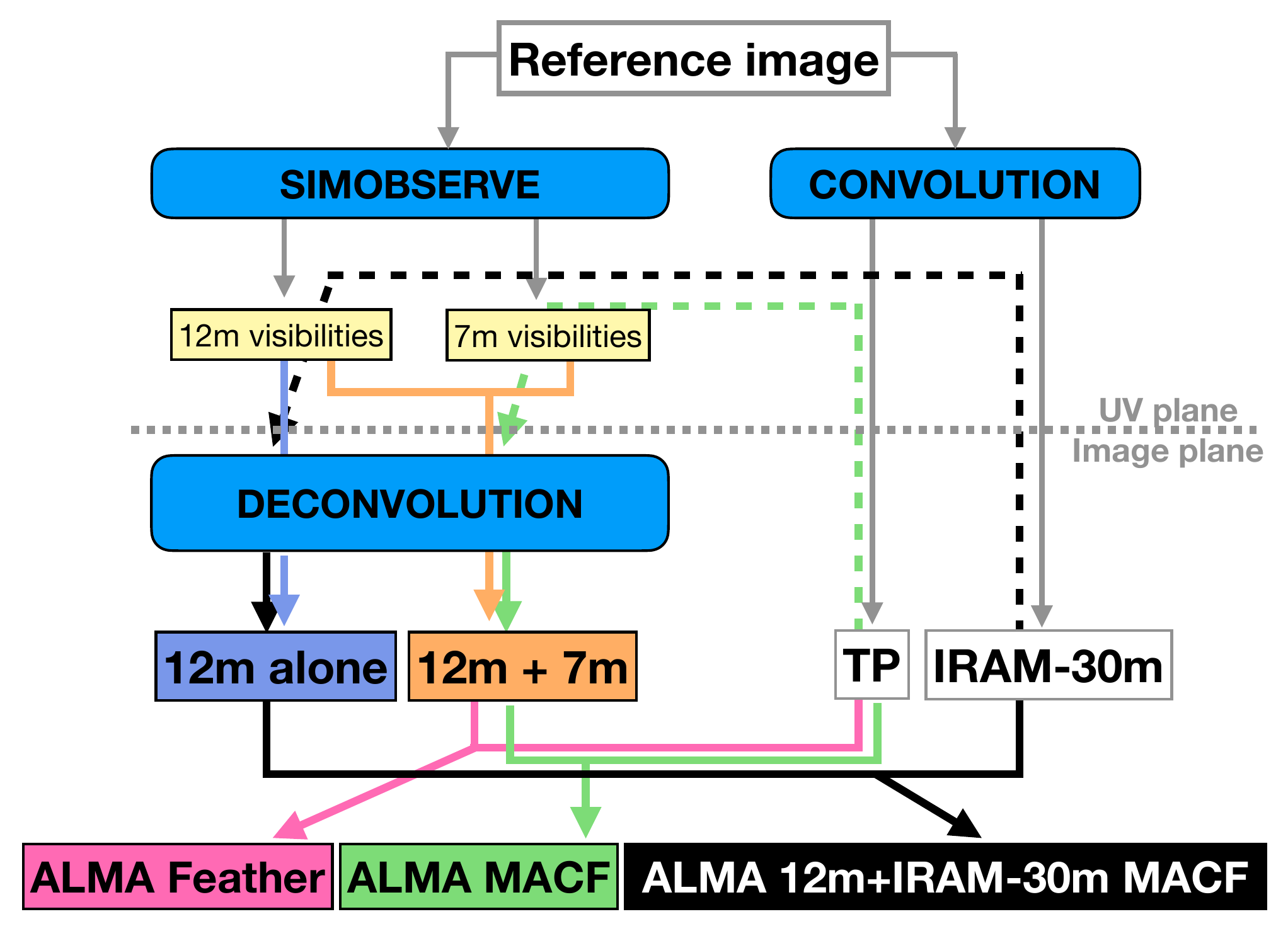}
	\caption{Schematic view of our simulation process.}
	\label{fig:data-combination}
\end{figure}

To quantify the impact of the spatial filtering effects on  different ISM structures
we create a series of synthetic observations of idealized Gaussian (cores) and cylinders (filaments) using the Common Astronomy Software Application \citep[CASA,][]{2022Casa}.
All these synthetic objects are placed at the center of a 6.25~arcmin wide field located at coordinates RA$=$12h00m00.0s and Dec$=$-23d00m00.0s, the optimal position in the sky that can be observed by ALMA in terms of source elevation and overall sensitivity (although not necessarily in terms of u-v coverage). 

We produce models characterized by a single frequency channel to avoid taking into consideration velocity effects. We select as central frequency 100 GHz, close to the frequency of the \nthp~molecular line, the most used dense gas tracers for cores and filaments \citep{2007Bergin} and representative of standard the continuum ALMA observations at 3mm (Band 3).  
We produce simulations of the three different ALMA arrays separately.
We perform our observations using the most compact configuration offered in ALMA 12m array, C43-1, providing an angular resolution $\theta_{res}(12m)$ of 3.38~arcsec and MRS $\theta_{MRS}(12m)$ of 28.5~arcsec. At these same frequencies, the ACA-7m array provides a $\theta_{res}(7m)$ and $\theta_{MRS}(7m)$ of 12.5~arcsec and 66.7~arcsec, respectively, while the TP antennas the $\theta_{res}(TP)$ goes down to 62.9~arcsec\footnote{All the values listed here are taken from the \textit{ALMA Cycle 10 Technical Handbook} \citep{2023Cortes}}. Since the primary beam (PB) of the ALMA antennas ($\sim$58~arcsec for a 12m antenna) is smaller than the full-width-half-maximum (FWHM) of the largest target (FWHM=100~arcsec), we observe our targets using mosaics.
We run our simulations in CASA v.6.5.2 \citep{2022Casa} using the task \texttt{simobserve} to simulate the visibility file and \texttt{tclean} for the imaging\footnote{All the information regarding the CASA tasks listed below are taken from https://casadocs.readthedocs.io/en/v6.5.2}. We adopt the same observational setup for all the synthetic datasets.

\subsection{Generation of visibilities with \texttt{simobserve}}\label{sec:simobs}

As first step in our simulations, and for each of our science targets, we generated their corresponding ALMA visibilities using the task \texttt{simobserve}. Taking a model image in FITS format as input, \texttt{simobserve} simulates the expected visibility dataset (MeasurementSet) obtained for a specific observational setup described by the following parameters:
\begin{itemize}
    \item \texttt{mapsize}, angular size of the mosaic map to simulate.
    \item \texttt{maptype}, position of the pointings for the mosaic observation.
    \item \texttt{pointingspacing}, spacing in between pointings.
    \item \texttt{integration}, integration time for each pointing.
    \item \texttt{totaltime}, total time of observation or number of repetitions.
\end{itemize}

For the 12m array simulations, we map the central 3.5~arcmin region of our fields as indicated in the \textit{ALMA Cycle 10 Technical Handbook} \citep{2023Cortes}. We cover this area using a 67-pointings mosaic following a Nyquist-sampled, hexagonal pointing pattern.
Each pointing in our mosaic is observed in two iterations (repetitions), 30 sec each, for a total of 1 min integration per pointing, typical for this type of ALMA observations.

For ACA-7m observations, the mapped area should be larger than the one observed with the 12m array by at least half of the PB, therefore we set it to 4.5~arcmin in size. We thus obtained a 33-pointings mosaic, selecting as integration time 30s and 14 repetitions. We repeat the observations for 3 consecutive days over transit mimicking a standard ALMA observing schedule.
Our choice of the integration time per pointing reproduces the expected time ratio C43-1 : ACA-7m = 1 : 7
following the official recommendations to match the sensitivity of the different ALMA arrays  \citep{2023Cortes}.

\subsection{Generation of images with \texttt{tclean}}

In a second step, we image each simulated visibility dataset using the CASA task \texttt{tclean}. This task is based on the CLEAN algorithm \citep{1974Hogbom}, the default method for obtaining an image from an interferometric observation, deconvolving it from the instrumental Point Spread Function (PSF). We set \texttt{mfs} as spectral definition mode (continuum imaging with only one output image channel), and select a standard Briggs weighting with a robust parameter equal to $0.5$. We set the pixel size as 0.5~arcsec and 2~arcsec ($\sim1/6 \cdot  \theta_{res}$) for the 12m and the ACA-7m arrays, respectively. To select the cleaning threshold, we produce first a dirty image from which we estimate its emission peak and set the threshold at 10$\%$ of the value and the number of maximum iterations at 10$^8$.
Afterwards we correct the images for the PB attenuation using the task \texttt{impbcor} setting a \texttt{cutoff} value of 0.85.
We independently image each of the 12m and the 7m arrays, as well as produce combined images using different combination techniques (see below).
Finally, we convolve the 12m and ACA-7m array images to obtain a symmetric synthesized beam using the task \texttt{imsmooth}. The final beam sizes are 4.5~arcsec and 16~arcsec for the 12m and ACA-7m array, respectively.

\subsection{Simulation of SD images with \texttt{imsmooth}}\label{sec:SDobs}

As third and last step, and in order to simulate SD observations to be combined with the ALMA 12m and ACA-7m array, we use the task \texttt{imsmooth}, which convolves the synthetic image at the required resolution. We produce simulations of TP observations at a resolution $\theta_{res}(TP)$ of 62.9~arcsec.

\subsection{Data combination methods}

To overcome the short-spacing problem we test different data combination methods already implemented in CASA \citep[see][and references therein]{2023Plunkett}. We show the workflow followed by our simulations in Fig.~\ref{fig:data-combination}. For convenience we briefly describe the different methods here.

First, we start combining the two interferometric visibilities (ALMA 12m + ACA-7m array) merging small and intermediate spatial scales. We use the CASA task \texttt{concat}, which concatenates several visibility datasets, applying weights of 1 and 0.116 to the simulated 12m and 7m visibilities, respectively, to take into account the different dish diameters \citep[this is only nedeed for simulated data, see][]{2023Plunkett}. We obtained a single interferometric MeasurementSet that we later imaged (joint-deconvolution) using \texttt{tclean}. We refer to this combined interferometric image as 12m + 7m, hereafter.

Secondly, we explore the Feather method \citep{2017Cotton} to combine the ALMA 12m + 7m interferometric datasets with the corresponding TP map. The Feather method is implemented in CASA as the \texttt{feather} task. This algorithm works in the image plane: it converts both interferometric (high-res) and SD (low-res) images to the visibility plane, combines them in the Fourier domain, and  transforms them back into the Feather image. We set the \texttt{sdfactor}, a parameter used to adjust the flux scale of the SD image, to 1.0 (usually constrained between 1.0 and 1.2). We refer to this image as ALMA Feather.

Thirdly, we also tested the Model-Assisted CLEAN plus Feather Method (MACF) \citep{2018Hacar,2023Plunkett}. This technique is a variant of Feather, which is actually used also here to combine the interferometric dataset with the SD image. As main difference, the MACF method first introduces the SD image as source model during the cleaning process of the interferometric visibilities (using the \texttt{startmodel} parameter), informing \texttt{tclean} about the location and brightness of the extended emission sampled by the SD \citep[for a more detailed discussion see][]{2023Plunkett}. We refer to this image as ALMA MACF.

\subsection{Source brightness}\label{sec:sourcebrightness}

The integrated sky brightness (i.e. flux density) $F_\nu$ in continuum is determined by the total dust column density $N(H_2)$ (proxied by the H$_2$ column density assuming an  gas-dust coupling) of the target source.
Following \citet{2008Kauffmann}, the conversion between the dust column density and flux density at a given frequency $\nu$ is described by 
\begin{equation}
    F_\nu=N(H_2) \Omega_A\ \mu_{H_2}\ m_H\ k_\nu B_\nu(T_d)
    \label{eq:flux}
\end{equation}
where $\mu_{H_2}=2.8$ is the mean molecular weight per hydrogen molecule, 
$\Omega_A=2.655\times10^{-11}\ sr\ (\frac{\Delta x_{pix}}{arcsec})^2$ the solid angle for a pixel size $\Delta x_{pix}=0.25$~arcsec in our models, and 
\begin{equation}
B_\nu(T_d)=1.475\times10^{-23} \frac{W}{m^2\ Hz\ sr} \left( \frac{\nu}{GHz}\right)^3\times \frac{1}{e^{0.0048(\frac{\nu}{GHz})(\frac{10\ K}{T_d})}-1}
\end{equation}
the Planck function given a dust temperature $T_{d}$. As for the dust emissivity, we adopted a standard $k_\nu=0.01$~cm$^{2}$g$^{-1}$ at 300~GHz (1~mm), which translates into $k_\nu=0.002$~cm$^{2}$g$^{-1}$ at 100~GHz (3~mm) assuming a frequency dependence $\nu^\beta$ with $\beta=1.6$ typical for filaments and cores in Orion \citep{Sadavoy2016,Mason2020}.

By exploring a series of simplified source geometries and array combinations (Sects.~\ref{sec:cores}-\ref{sec:iram}) we aim to isolate the effects of the interferometer response and quantify its effects on the quality of the observations. 
In order to simulate realistic ISM conditions, we have adopted a source peak intensity of $I_0=7.5\times 10^{-3}\ mJy/pix$ in all our synthetic models.
According to Eq.~\ref{eq:flux}, this flux density corresponds to an equivalent dust column density of $N(H_2)\sim 1.5\times 10^{23}$~cm$^{-2}$ (or A$_V\sim$~150 mag) at $\nu=100$~GHz and assuming a constant $T_{dust}=$~15~K, similar to the maximum peak column densities found in cores and dense filaments in active star-forming regions such as Orion \citep[e.g.,][]{2018Hacar}. While undoubtedly bright, these high peak fluxes guarantee a bright emission in our synthetic observations in order to investigate the impact of different interferometric filtering effects.

\subsection{Thermal noise}\label{sec:thermalnoise}

To clearly identify interferometric artefacts such as sidelobes it is convenient to run most of our simulations without adding thermal noise (noise-free). 
Yet, and to investigate these effects under realistic observing conditions, we also produce (noisy) simulations including atmospheric thermal noise assuming standard weather conditions for Band 3 with precipitated water vapor (PWV) of 1.8~mm \citep[5th PWV octile according to the ALMA Handbook;][]{2023Cortes}. We include noise in our ALMA simulations (Sect.~\ref{sec:simobs}) by setting the parameter \texttt{thermalnoise} as \textit{tsys-atm} in the task \texttt{simobserve}. Likewise, we add an additional thermal noise to our SD data using the radiometer formula with similar $T_{sys}$ as the interferometric data and the corresponding integration times (see Sect.~\ref{sec:iram}).

\section{Observing isolated cores: single Gaussian analysis}\label{sec:cores}

\begin{figure*}[htbp]
	\centering
	\includegraphics[clip=true,trim=0.0cm 0.0cm 0.0cm 0.0cm,width=0.9\textwidth]{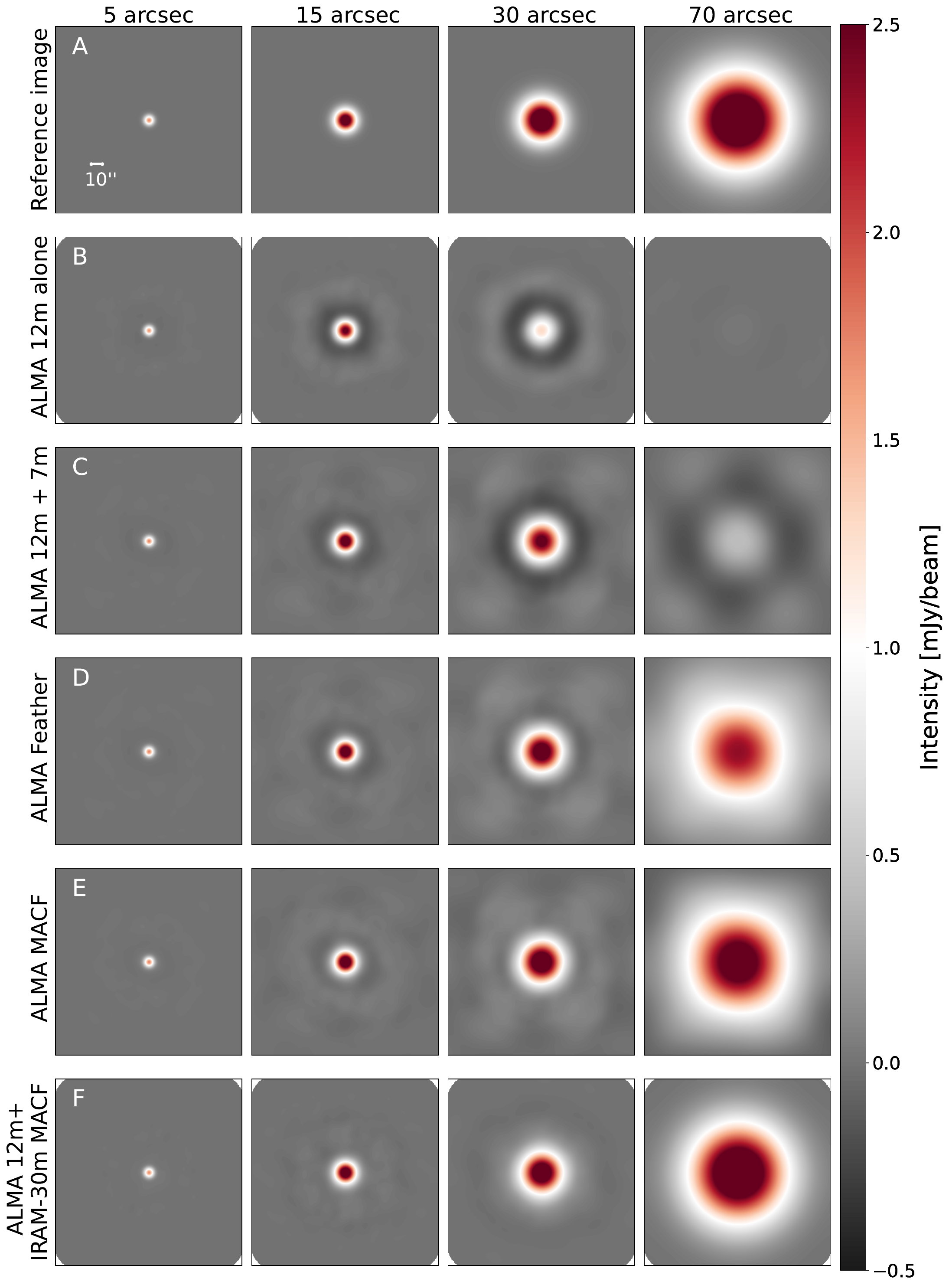}
	\caption{\textbf{From left to right:} Noise-free simulations of isolated cores with FWHM$_0$=5, 15, 30 and 70~arcsec. The first row shows the synthetic reference images used as input for the simulations (panel A). The rows below display the results for different methods of data combination. \textbf{From top to bottom}: ALMA 12m alone (panel B), ALMA 12m + 7m (panel C), ALMA Feather (panel D), ALMA MACF (panel E), and ALMA 12m + IRAM-30m MACF (panel F). The size of each image is 2.8~arcmin $\times$ 2.8~arcmin.}
	\label{fig:cores_comparison}
\end{figure*}

\begin{figure*}[htbp]
	\centering
	\includegraphics[clip=true,trim=0.0cm 0.0cm 0.0cm 0.0cm,width=\textwidth]{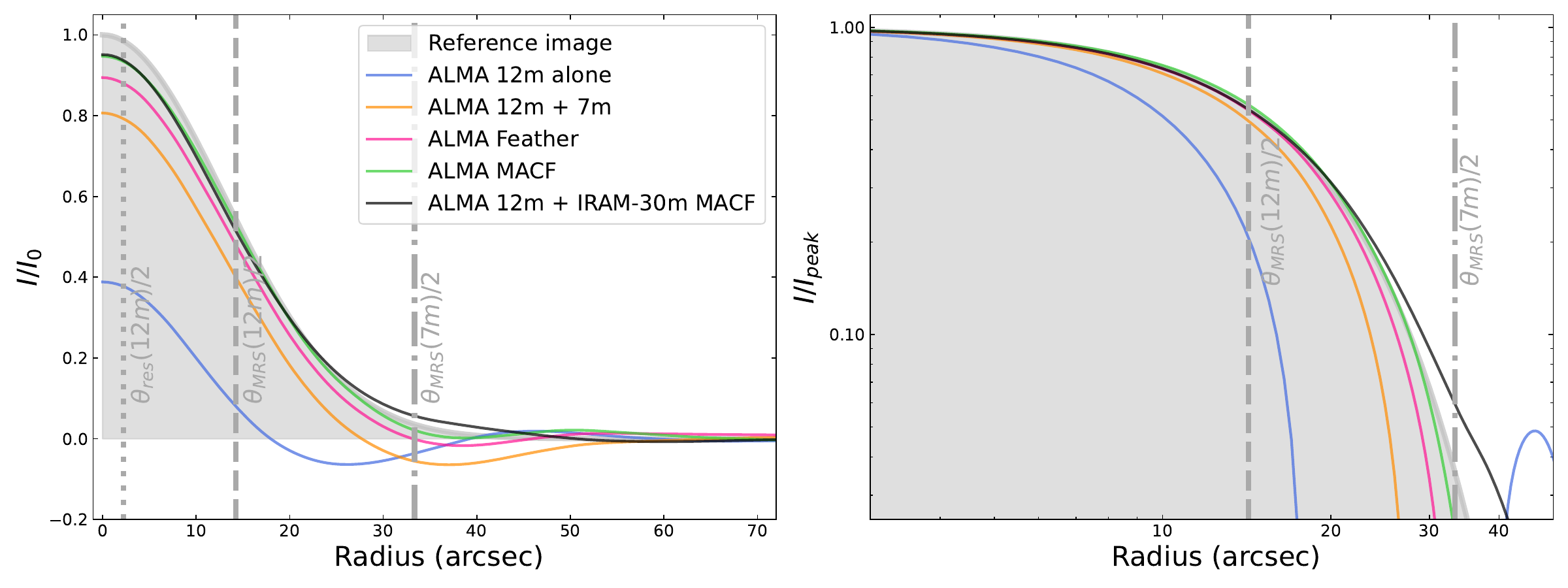}
	\caption{\textbf{Left panel}. Radial profile extracted from a horizontal cut in the center of a FWHM=30~arcsec Gaussian core image normalized with respect to the reference image peak (shown by the grey shaded area) displayed in linear scale. \textbf{Right panel}. The same radial profile normalized with respect to its peak value (the reference profile is shown by the grey shaded area) displayed in log-scale. The data combination methods used are marked in different colors: ALMA 12m alone (blue), ALMA 12m + 7m (orange), ALMA Feather (pink), ALMA MACF (green) and ALMA 12m + IRAM-30m MACF (black).}
	\label{fig:cores_30arcsec_profile}
\end{figure*}

Dense cores are the sites of stellar birth \citep{1989Benson}. They can be identified as compact objects within molecular clouds in (sub-)millimetre continuum emission \citep{2014Andre}. Cores are characterized by a roundish shape with a typical diameter of 0.03-0.2 pc \citep{2007Bergin}, that is $\sim$13-90~arcsec at the Orion distance \citep[$\sim$~420~pc; see][]{2007Menten}. Resolved at the SD resolution in nearby clouds such as Taurus ($\sim$~140~pc), dense cores are routinely studied in continuum showing typical radii of 20-75 arcsec using SD bolometers \citep[e.g.,][]{2002Tafalla}. 
Cores exemplify the simplest geometry to be recovered by the interferometer and so are the first targets in our study.

Despite their favourable conditions, previous ALMA observations of nearby star-forming regions reveal the difficulties of interferometers recovering the full emission profiles of dense cores. Continuum interferometric-alone observations in local molecular clouds ($\sim$130-200~pc) such as Chamaleon \citep[Cycle 1, 12m-alone;][]{2016Dunham}, Ophiuchus \citep[Cycle 2, 12m-alone][]{2017Kirk}, or Taurus \citep[Cycle 4-6, ACA 7m-alone][]{2020Tokuda} result into few core detections with compact radii of few times the $\theta_{res}$ ($\sim$5~arcsec for the 12m data and $~\sim15-20$~arcsec in the case of 7m data).  These core radii are significantly smaller than the expected (deconvolved) sizes estimated from the corresponding SD maps in these fields \citep[see Fig.4 in][]{2017Kirk}. Despite the high sensitivity of their maps, most of these studies also report a high number of non-detections \citep[e.g., 54 non-detections out of the 73 fields observed by][]{2016Dunham}.    

In contrast, the combination of 12m+7m data shows a significant improvement of the amount of extended emission recovered \citep[e.g.,][]{Tokuda2016} and a systematically increased core radius in dedicated studies of centrally condensed cores in Taurus \citep[see Fig.1 in ][for an example]{2019Caselli}. Similar improvements after data combination are also seen in ALMA surveys of dense cores in Orion despite their smaller angular size \citep[e.g.,][]{Ohashi2018,2020Dutta}. Characterizing the effects of interferometric filtering is therefore of paramount importance to assess the accuracy of the core properties (mass, size, and column density) derived from these observations.

The column density profiles of cores have been described using different power-law models \citep{2001Alves,2002Tafalla}. For simplicity, however, we decided to model our cores as 2D-Gaussians of different sizes (defined by their full width half maximum, FWHM) described by
\begin{equation}
    G(x,y)=I_0 \exp{\left(-\frac{(x-x_0)^2}{2\sigma_x^2}-\frac{(y-y_0)^2}{2\sigma_y^2}\right)},
    \label{eq:core}
\end{equation}
where \textit{I$_0$} corresponds to the core peak intensity, $x_0$ and $y_0$ its central coordinates, and $\sigma_x,\sigma_y$ its two spatial dispersions which can be expressed in terms of the core full-width-half-maximum (FWHM) as $\sigma_x=\sigma_y=\frac{FWHM}{2\sqrt{2\ln{2}}}$ . 
We produced a sample of several different synthetic cores with FWHM$=$5, 10, 15, 20, 30, 40, 50, and 70~arcsec.

We produce noise-free ALMA simulations of all our cores using different arrays and combinations. Fig.~\ref{fig:cores_comparison} shows the results for cores with different increasing FWHM (form left to right) of 5~arcsec (similar to $\theta_{res}(12m)$), 15~arcsec, 30~arcsec ($\sim \theta_{MRS}(12m)$) and 70~arcsec, and different combinations (from top to bottom) all convolved to the same resolution and shown within the same flux range to facilitate their comparison. The top row (panel A) displays the synthetic reference images used as input. 

We show the ALMA 12m C43-1 alone simulations on the second row of Fig.~\ref{fig:cores_comparison} (panel B). The interferometer filtering effect is clearly visible in these images as the core size grows. At 5~arcsec ($\sim\theta_{res}(12m)$, first panel) the interferometer is able to reproduce the source with high fidelity. However, as the FWHM of the core increases, the interferometer is not able to recover the total emission of the source anymore (the object becomes fainter) and, in addition, negative sidelobes surrounding the emission become progressively more and more prominent. Interestingly, the flux losses and negative sidelobes become already apparent at FWHM $\sim$15~arcsec, that is at scales $<\theta_{MRS}(12m)$. We note that the most significant effect is seen at 70~arcsec, where the core is no longer visible due to the heavy filtering at large scales.

The recovery of the shape and flux of the emitting source improves with data combination. We show our ALMA 12m + 7m array simulations on the third row of Fig.~\ref{fig:cores_comparison} (panel C). The overall fraction of emission recovered at intermediate angular scales increases for FWHM values between 15-30~arcsec. Also, a faint emission is now visible in case of FWHM=70~arcsec.

However, we notice that the negative sidelobes are still prominent and present a less-circular pattern due to the addition of the 7m data.
The ALMA TP contribution is added in the lower rows of Fig.~\ref{fig:cores_comparison}. We show the ALMA Feather and ALMA MACF simulations in panels D and E rows, respectively. The large scale sensitivity allows to recover the emission even at 70~arcsec (last column) as to reduce the image artifacts stemming from PSF sidelobes.

As representative example of the observed core behavior, we display the radial profile extracted from a horizontal cut in the center of a FWHM=30~arcsec core in Fig.~\ref{fig:cores_30arcsec_profile}. On the left panel, we show the profile in linear scale compared to the reference (input) peak value ($I_0$), while on the right in logarithmic scale with values relative to the peak of each profile ($I_{peak}$). As shown by the ALMA 12m alone profile (blue line), the value recovered at the center is less than 50$\%$ of the one in the reference (left panel). The radial profile of the ALMA 12m alone data also decreases faster than the one in the reference (see also right panel) and shows clear negative sidelobes at around 25~arcsec, i.e. it is not a scaled-down version of the reference profile. On the other hand, the ALMA 12m + 7m array (orange line) recovers higher flux values ($\sim80\%$) and accurately reproduces the shape of the profile up to $\sim15$~arcsec, beyond it slightly diverges decreasing faster than the reference one. 
Compared to the interferometric images, addition of the SD data in the the ALMA Feather and MACF (pink and green lines, respectively) significantly improves the flux recovery $\geq90\%$ across the entire profile with only small ($<10\%$) deviations at large radii.
The quantitative comparison of the different radial profiles illustrates how the lack of the short-spacinginformation could potentially affect the emission properties recovered by interferometers, such as ALMA, even for simple sources such as our idealized Gaussian cores.

\subsection{Quantifying the effects of the short-spacing information in simple geometries}

\begin{figure}[htbp]
	\centering
	\includegraphics[clip=true,trim=0.0cm 0.0cm 0.0cm 0.0cm,width=\columnwidth]{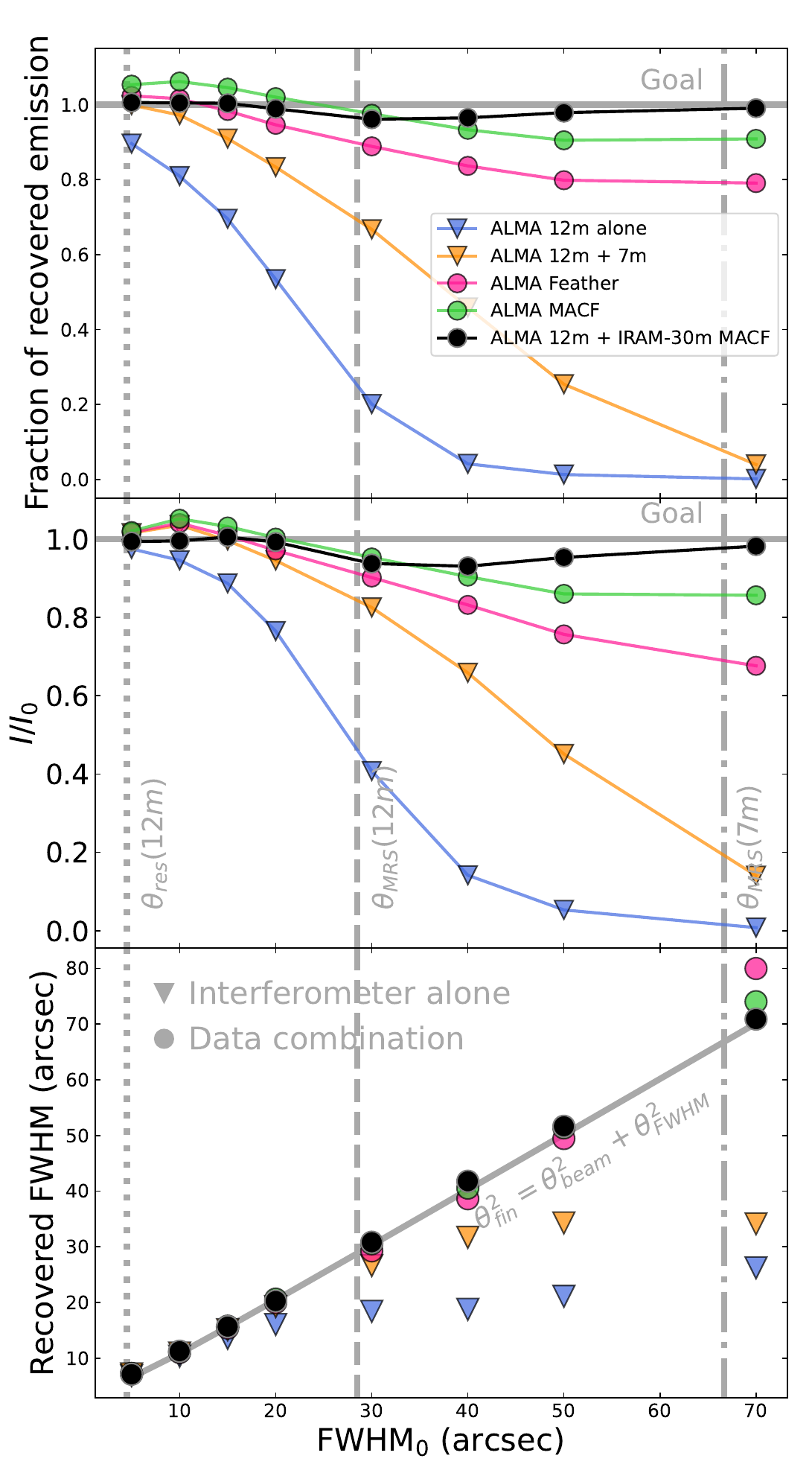}
	\caption{Reconstructed properties of isolated Gaussian cores. 
		From top to bottom:
		\textbf{(Top panel)} Percentage of emission recovered along the entire radial profile vs. the FWHM$_0$. Different colors are used for the different data combination methods used: blue for the ALMA 12m alone, orange for the 12m + 7m, pink for the ALMA Feather, green for the ALMA MACF, and black for the ALMA 12m + IRAM-30m MACF. The grey dotted and dashed lines show $\theta_{res}(12m)$ and $\theta_{MRS}(12m)$, respectively. 
		\textbf{(Mid panel)}. Intensity peak estimated from the Gaussian fit vs. FWHM$_0$. 
		\textbf{(Lower panel)}. FWHM estimated from the Gaussian fit vs. FWHM$_0$. The grey solid line shows the theoretical expectation. For comparison, the results for the radial profile shown in Fig.~\ref{fig:cores_30arcsec_profile} with FWHM$_0$=30 arcsec correspond to the values close to the $\theta_{MRS}(12m)$.}
	\label{fig:cores_properties}
\end{figure}

We further quantify the impact of the intrinsic spatial filtering of the interferometer and of data combination on the recovery of the true properties of the emitting source based on the characterization of the above radial profiles, this time for all our cores with different angular sizes. In Fig.~\ref{fig:cores_properties}, we measure the fraction of recovered flux (\textit{top panel})), peak flux (I; \textit{mid panel}), final full-width-half-maximum (FWHM; \textit{lower panel})) with respect to the input model (reference) for the different interferometric (inverted triangles) and combined (circles) reductions.

We first compare the fraction of total flux recovered by our different reductions in Fig.~\ref{fig:cores_properties} (\textit{top panel}) as function of the original core FWHM$_0$.
Even for objects with a FWHM$_0\sim \theta_{res}(12m)$ (grey dotted line), the ALMA 12m alone (blue) is not able to recover the total emission of the source (already $\sim90\%$) and, as the size of the object grows, the fraction of recovered emission decreases dramatically. Around the $\theta_{MRS}(12m)$ (grey dashed line in Fig.~\ref{fig:cores_properties}), the ALMA 12m alone is only able to recover only up to 20$\%$ of the emission, a smaller value compared to the expected $1/e$ definition by \citet{1994Wilner}). The ALMA 12m alone is clearly affected by the spatial filtering effect, in particular by the lack short-spacing information, not being able to recover emission at large angular scale (half of the total emission lost in cores with FWHM~$\sim$20~arcsec $<\theta_{MRS}(12m)$).

The ALMA 12m + 7m profile (in orange in Fig.~\ref{fig:cores_properties}) shows a similar decreasing trend, although much shallower. Adding data sensitive to intermediate-scales contributes to recover a much larger fraction of the core emission ($\gtrsim$70$\%$ up to sizes of $\sim\theta_{MRS}(12m)$). Still, the flux recovery continues to decrease for cores with larger radii down to scales comparable to $\theta_{MRS}(7m)$, where ALMA 12m + 7m array is losing up to 90$\%$ of the core emission.
Combining ALMA 12m plus 7m data has improved the fraction of emission recovered especially at intermediate scales, but interferometers alone are still affected by significant spatial filtering effects beyond 30~arcsec.

In contrast, the addition of the SD information shown by ALMA Feather and MACF profiles (displayed in pink and green in Fig.~\ref{fig:cores_properties}, respectively) always recovers a fraction $\ge 80 \%$ independent of the core size. Both profiles overestimate the fraction of recovered emission at small scales (FWHM$\le$20~arcsec) up to a 10$\%$ factor and show a decreasing trend towards larger sizes.
Overall, the addition of large scale contributions improves the recovery of the emission of objects at all scales (up to 90$\%$ at $\theta_{MRS}(12m)$), allowing to estimate the true flux of the sources within a $\sim20\%$ uncertainty. The ALMA Feather and MACF show similar but not identical profiles. However their differences are always below 10$\%$, a level of discrepancy that would probably not be detected if the observational and instrumental noise would be added.

We further fit the resulting radial profile with a Gaussian function to estimate the intensity peak and the FWHM, similar to observations. We  show the normalized intensity peak recovered for cores of different sizes in
Fig.~\ref{fig:cores_properties} (\textit{mid panel}). The behaviour of different methods of data combination is similar to the recovered emission along the entire profile. 
We estimate the peak values recovered by the ALMA 12m alone (blue) at $\theta_{MRS}(12m)$ to be around 40$\%$ ($>20\%$, the fraction of recovered emission at the same scale). The two results are compatible since the fraction of recovered emission takes into account also the negative sidelobes that are not included in the fitting process, resulting in lower values when accounting for the emission along the entire profile. However, even if the effects are less severe for the intensity peak, short-spacing issues are still present and visible, especially at large scales.
Similar trends are shown by the ALMA 12m + 7m (orange), the ALMA Feather (pink) and MACF (green) with respect to the fraction of recovered emission. The lack of short-spacing information remains noticeable in our interferometric-alone (12m + 7m) dataset, and it is only mitigated by the combination with the ALMA TP (ALMA Feather and MACF).

To investigate the effects on the recovered source size, we also display FWHM estimated from the Gaussian fit in respect to the reference value FWHM$_0$ in Fig.~\ref{fig:cores_properties} (\textit{lower panel}). In case of a perfect recovery, the observed FWHM of our cores is expected to follow the initial FWHM$_0$ convolved with the beam resolution (added in quadrature), as shown by the grey solid line in Fig.~\ref{fig:cores_properties}. While the values recovered by the ALMA 12m alone (blue) for marginally resolved cores with FWHM$_0\sim$5-10~arcsec follow the expected theoretical prediction, the observed FWHM is systematically underestimated for larger FWHM$_0$ and saturates to a constant value at  FWHM$_0\ge$30~arcsec.
In the case of cores with sizes on the order of $\theta_{MRS}(12m)$, the estimated FWHM is around 15~arcsec, almost a factor of 2 lower than FWHM$_0$. As shown by the ALMA 12m alone profile, the spatial filtering is not only acting on the flux recovery, but also on the estimate of the object FWHM. The lack of intermediate and large scale observations is critically affecting the recovered FWHM, systematically producing smaller cores.

In comparison, the ALMA 12m + 7m results (marked in orange in Fig.~\ref{fig:cores_properties}) are in better agreement with the theoretical predictions up to FWHM$_0\sim$30~arcsec. Still, this method again underestimates the FWHM of cores with FWHM$_0\gtrsim$40~arcsec, up to a factor of 50\% at 70~arcsec (the 7m array $\theta_{MRS}(7m)$), limited again by the filtering effect at large scales. As already seen for the fraction of recovered emission, the addition of short-spacing information
in the ALMA Feather and ALMA MACF methods (pink and green dots, respectively) allows a better estimate of the true FWHM. The resulting FWHM values closely reproduce the expected theoretical growth (grey line) within $\sim5\%$ error.

\subsection{Additional biases in realistic interferometric observations}\label{sec:corenoise}

\begin{figure*}[ht!]
	\centering
	\includegraphics[width=\textwidth]{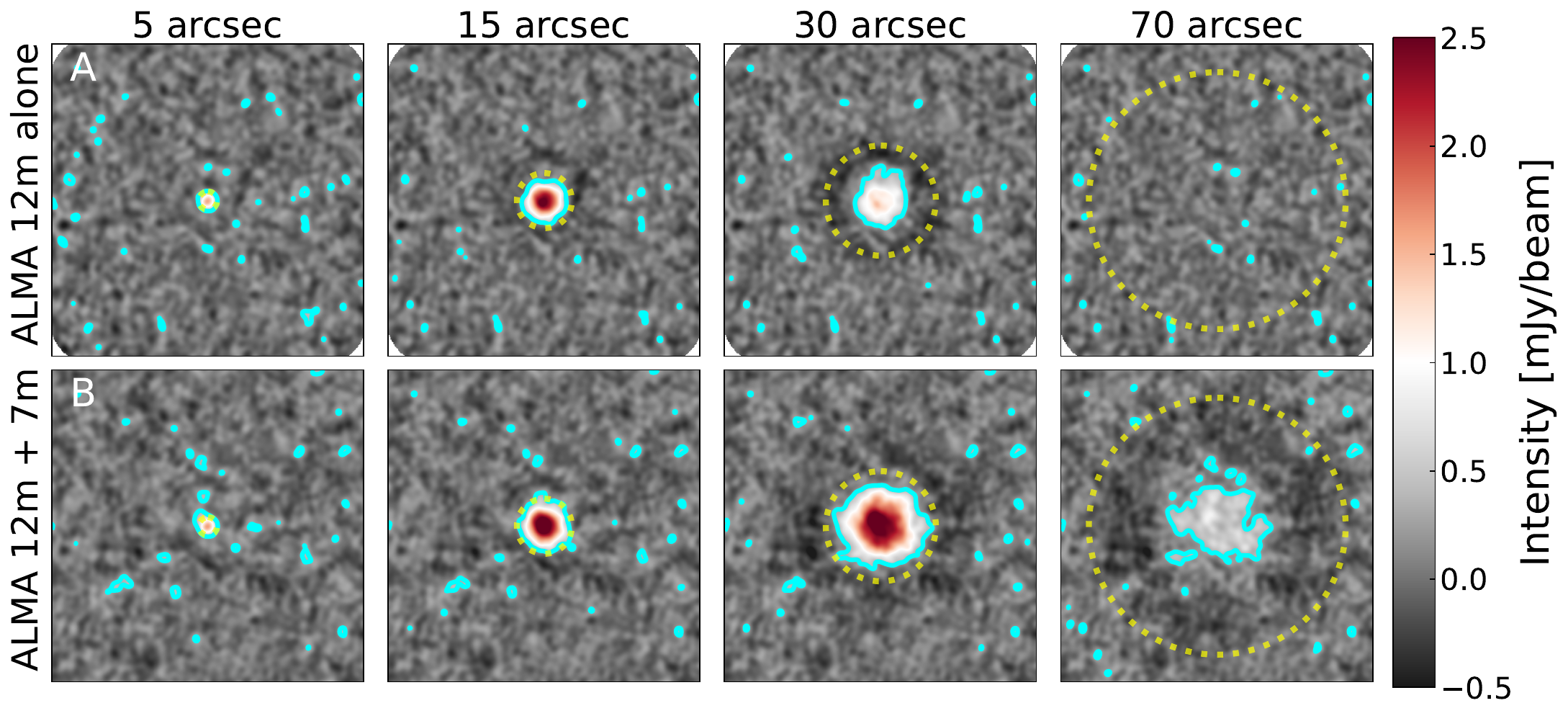}
	\caption{Interferometric ALMA 12m alone (panel A) and 12m+7m (panel B) simulations of isolated cores with FWHM$_0$=5, 15, 30 and 70~arcsec (from left to right) including thermal noise with PWV=1.8~mm. In each image, we highlight the emission $\geq 3\sigma$ (cyan contours) compared to the core FWHM$_0$ (yellow dotted circles).}
	\label{fig:noisedetection}
\end{figure*}

The above issues found in interferometric-only datasets could potentially hamper the detection of cores in real observations.
Figure~\ref{fig:noisedetection} shows the recovered emission of cores with FWHM~=~5, 15, 30, and 70 arcsec in our ALMA 12m alone (panel A) and 12m+7m (panel B) datasets, this time using our noisy simulations with PWV=1.8~mm (see Sect.~\ref{sec:thermalnoise}). 
The reduction of the peak intensity in cores of increasing FWHM in interferometric-only observations (see Fig.~\ref{fig:cores_properties}) effectively reduces their signal-to-noise ratio (S/N). 
Eventually buried within the noise, these effects can severely affect the completeness of dense core surveys biasing their detection rates towards compact, unresolved sources (i.e. cores with FWHM$_0\lesssim \theta_{MRS}(12m)$). 

More subtle, the observational signatures of the interferometric filtering could go unnoticed in real observations. While clearly visible in our previous noise-free images (see panels B and C in Fig.~\ref{fig:cores_comparison} for comparison) negative sidelobes can also be hidden within noisy images giving the false impression of a non-detection in a high sensitivity dataset. A comparison between the real core radius (yellow dotted circles) and the emission contour corresponding with a S/N=3 in our images (cyan contours) demonstrate how hidden negative emission features can inadvertently alter the recovered FWHM of partially resolved sources if these effects were ignored.   

Despite their simplicity, our synthetic experiments illustrate how spatial filtering effects can critically bias the derived core properties in case of interferometric-alone observations.
ALMA 12m alone datasets are expected to produce artificially low-intensity, narrow cores in continuum (Fig.~\ref{fig:cores_properties}) explaining many of the observational properties and non-detections obtained in previous ALMA studies (see above). Only the addition of the {\it short-spacing} information (at least 12m+7m although ideally 12m+7m+TP) can guarantee a reliable detection rate and estimate of the actual core emission properties (i.e. total flux, peak flux, and FWHM) with an accuracy better than 10\% \citep[e.g., see][]{2019Caselli}. 

The above interferometric biases can be particularly severe in the case of ALMA observations of resolved targets in nearby clouds \citep[e.g.,][]{2016Dunham}.
However, similar effects are expected in more distant targets when observed with extended ALMA configurations if $FWHM_0>\theta_{MRS}(12m)$. These observational biases should be considered when characterizing physical properties derived from the observed fluxes such as the core mass, peak column density, and size.
Given the broad range of column densities and sizes reported for cores in nearby star-forming regions \citep[e.g.,][]{2020Konyves}, a careful consideration of these biases is essential when deriving statistical distributions such as the Core Mass Function using interferometric observations (see Appendix~\ref{sec:cmf} for a discussion). In case of doubt, observers should use the full ALMA 12m+7m+TP array capabilities when targeting these sources.



\section{Observing isolated filaments: effects on elongated geometries}\label{sec:fil}

\herschel~far-IR surveys \citep{2010Andre,2010Molinari} have revealed the presence of a network of filamentary structures permeating the ISM. The analysis of recent continuum maps, provides the first systematic and homogeneous measurements of key physical properties such as the filament mass, radial profile, and width \citep{2011Arzoumanian, 2019Arzoumanian, 2013Palmeirim, 2015Konyves}. Using \herschel~observations down to 18~arcsec resolutions, filaments appear to be described by a characteristic typical width of $\sim$0.1 pc \citep{2011Arzoumanian, 2019Arzoumanian}.
However, molecular line ALMA observations at resolutions of 4.5~arcsec unravelled an unexpected physical and kinetic complexity of filamentary networks of the ISM \citep[e.g.,][]{2013Peretto,2018Hacar,2019Shimajiri,Chen2019}. Following \citet{2018Hacar}, the filamentary structure identified by \herschel~in the two Orion Molecular Clouds OMC-1 and OMC-2 appears to be a collection of small-scale filaments, the so-called fibers, characterized by a narrower width of $\sim$0.03~pc detected using ALMA observations of dense tracers such as \nthp. Since interferometric resolutions are needed to resolve filaments up to $\sim$0.03 pc (or 14~arcsec at the distance of Orion), we aim to quantify in this section the impact on the spatial filtering effects on the analysis of these fine gas sub-structures.

\subsection{Single Gaussian filament analysis}\label{sec:gaussianfils}

\begin{figure*}[htbp!]
	\centering
	\includegraphics[clip=true,trim=0.0cm 0.0cm 0.0cm 0.0cm,width=0.85\textwidth]{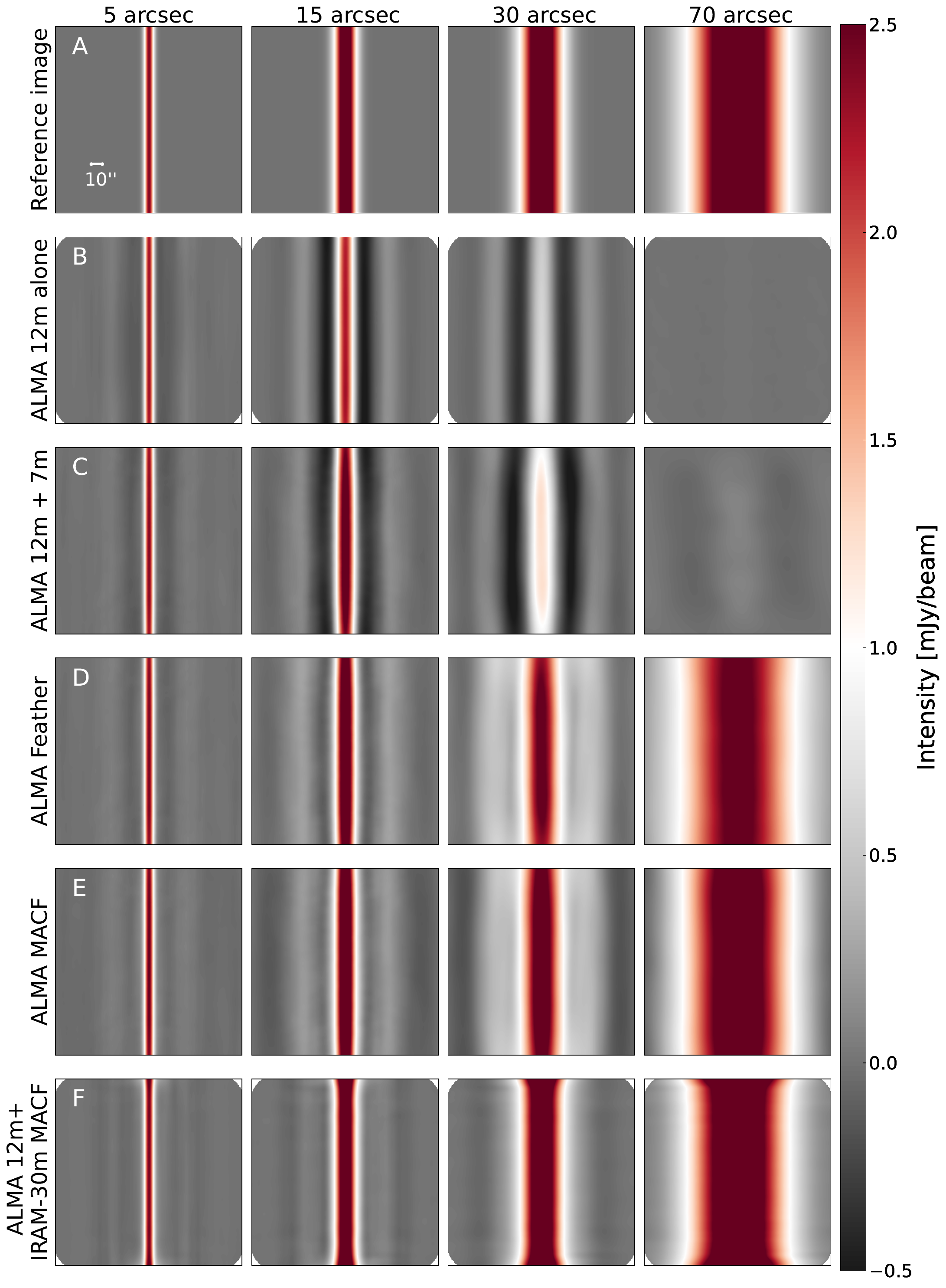}
	\caption{\textbf{From left to right:} Noise-free simulations of isolated Gaussian filaments with FWHM$_0$=5, 15, 30 and 70~arcsec. The first row shows the synthetic reference images used as input for the simulations (panel A). The rows below display the results for different methods of data combination. \textbf{From top to bottom}: ALMA 12m alone (panel B), ALMA 12m + 7m (panel C), ALMA Feather (panel D), ALMA MACF (panel E), and ALMA 12m + IRAM-30m MACF (panel F). The size of each image is 2.8~arcmin $\times$ 2.8~arcmin.}
	\label{fig:gausfil_comparison}
\end{figure*}

\begin{figure*}[htbp]
	\centering
	\includegraphics[clip=true,trim=0.0cm 0.0cm 0.0cm 0.0cm,width=\textwidth]{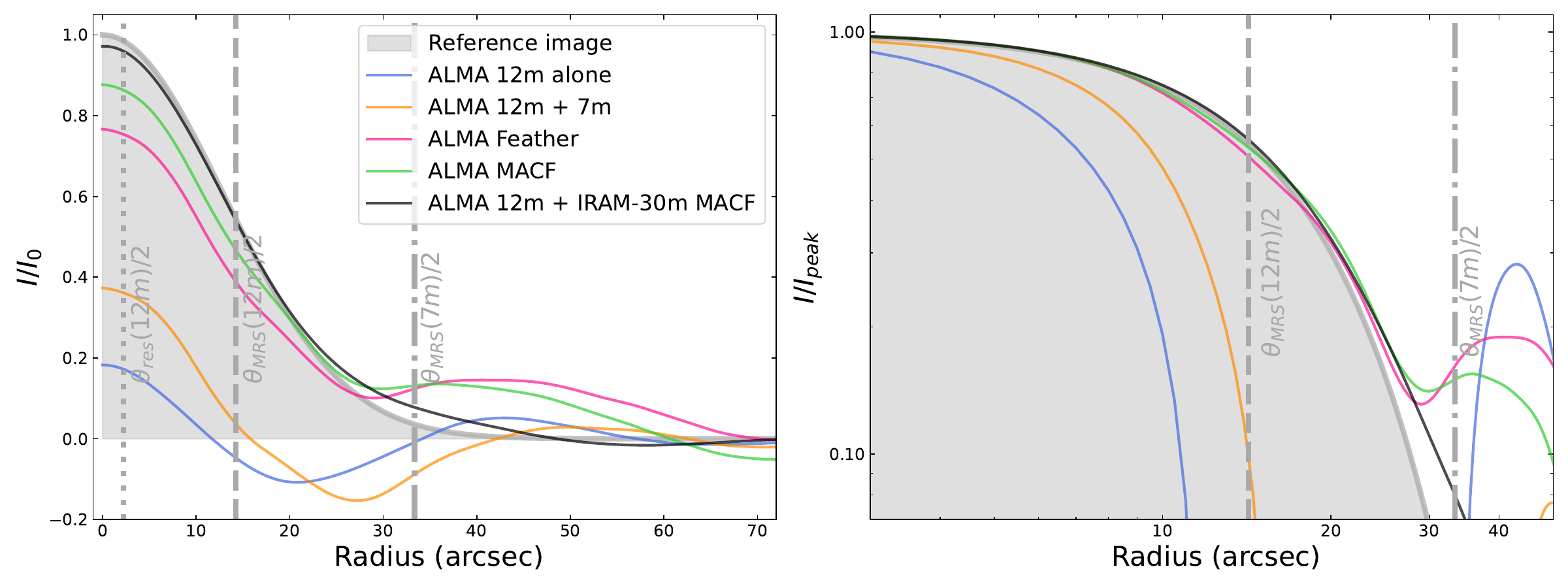}
	\caption{\textbf{(Left panel)} Radial profile extracted through a horizontal cut in the center of a FWHM$_0$=30~arcsec Gaussian filament image normalized with respect to the reference image peak (shown by the grey shaded area) displayed in linear scale. \textbf{(Right panel)} Same radial profile this time normalized with respect to the corresponding observed peak value in each map $I_{peak}$ displayed in log-scale. In both panels the reference profile is shown by the grey shaded area.
 The results of the distinct data combination methods are indicated with different colors: ALMA 12m alone (blue), ALMA 12m + 7m (orange), ALMA Feather (pink), ALMA MACF (green) and ALMA 12m + IRAM-30m MACF (black).}
	\label{fig:gausfil_30arcsec_profile}
\end{figure*}

For simplicity, we first simulate the observation of a simple object elongated towards one direction. We generate  noise-free, synthetic observations of an infinitely long filament with a 1D-Gaussian radial profile along the x-axis (similar to the profile of the core, see Sec.~\ref{sec:cores})
\begin{equation}
    G(x)=I_0\ \exp{\left(-\frac{(x-x_0)^2}{2\sigma^2}\right)}
\end{equation}
with \textit{I$_0$} as the peak intensity, $x_0$ position of the peak, x the impact parameter and $\sigma=\frac{FWHM_0}{2\sqrt{2 \ln{2}}}$, and a constant profile along the y-axis. We modeled Gaussian filaments with FWHM$=$5, 10, 15, 20, 30, 40, 50 and 70~arcsec. 

In Fig.~\ref{fig:gausfil_comparison}, we show the results of four representative filaments with FWHM of 5~arcsec (similar to $\theta_{res}(12m)$), 15~arcsec, 30~arcsec ($\sim \theta_{MRS}(12m)$) and 70~arcsec (panel A) observed using different interferometric observations (panels B-C) and data combination methods (panels D-E), all display within the same intensity range. 
Qualitatively, the results are similar to those found for cores (Fig.~\ref{fig:cores_comparison}) in which the inclusion of first 7m data (panel C) and later TP observations (e.g. ALMA Feather) systematically improves the recovery of the true sky emission with respect to the 12m alone simulations (panel B). Compared to our previous core simulations, however, the simulation of filamentary structures appear to produce larger filtering effects and more prominent sidelobes.

We also display the radial profile extracted from an horizontal cut in the center of a FWHM=30~arcsec Gaussian filament in Fig.~\ref{fig:gausfil_30arcsec_profile} both in linear scale and absolute units (left panel) as well as in relative units and in logarithmic scale with respect to the peak of each profile (right panel).
The results obtained for elongated structures are noticeably more dramatic with respect to cores of similar FWHM (see Fig~\ref{fig:cores_30arcsec_profile}). 
As shown by the ALMA 12m alone profile (blue line in Fig.~\ref{fig:gausfil_30arcsec_profile}), the recovered peak is only around 20\% (a factor of 2 lower than in the cores) and there is a negative sidelobe of comparable intensity to the positive emission.
The ALMA 12m + 7m array (orange line in Fig.~\ref{fig:cores_30arcsec_profile}) is recovering 40$\%$ of the peak value (80$\%$ for the core), still showing a negative sidelobe around 30~arcsec. Only the ALMA Feather and MACF (pink and green lines in Fig.~\ref{fig:gausfil_30arcsec_profile}, respectively) profiles are closer to the reference one although not even these combination methods are able to recover the 100$\%$ of the peak flux in filaments. The ALMA Feather and MACF profiles also show deviations from the reference Gaussian profile at large radii (clearly visible around 35~arcsec in the \textit{right panel} of Fig.~\ref{fig:gausfil_30arcsec_profile}). The more severe filtering effects seen on these filaments with respect to Gaussians of similar FWHM suggest that the long dimension in elongated emission features introduce additional artefacts affecting the flux recovery at all scales.

\subsection{Effects on elongated geometries}

\begin{figure}[htbp]
	\centering
	\includegraphics[clip=true,trim=0.0cm 0.0cm 0.0cm 0.0cm,width=\columnwidth]{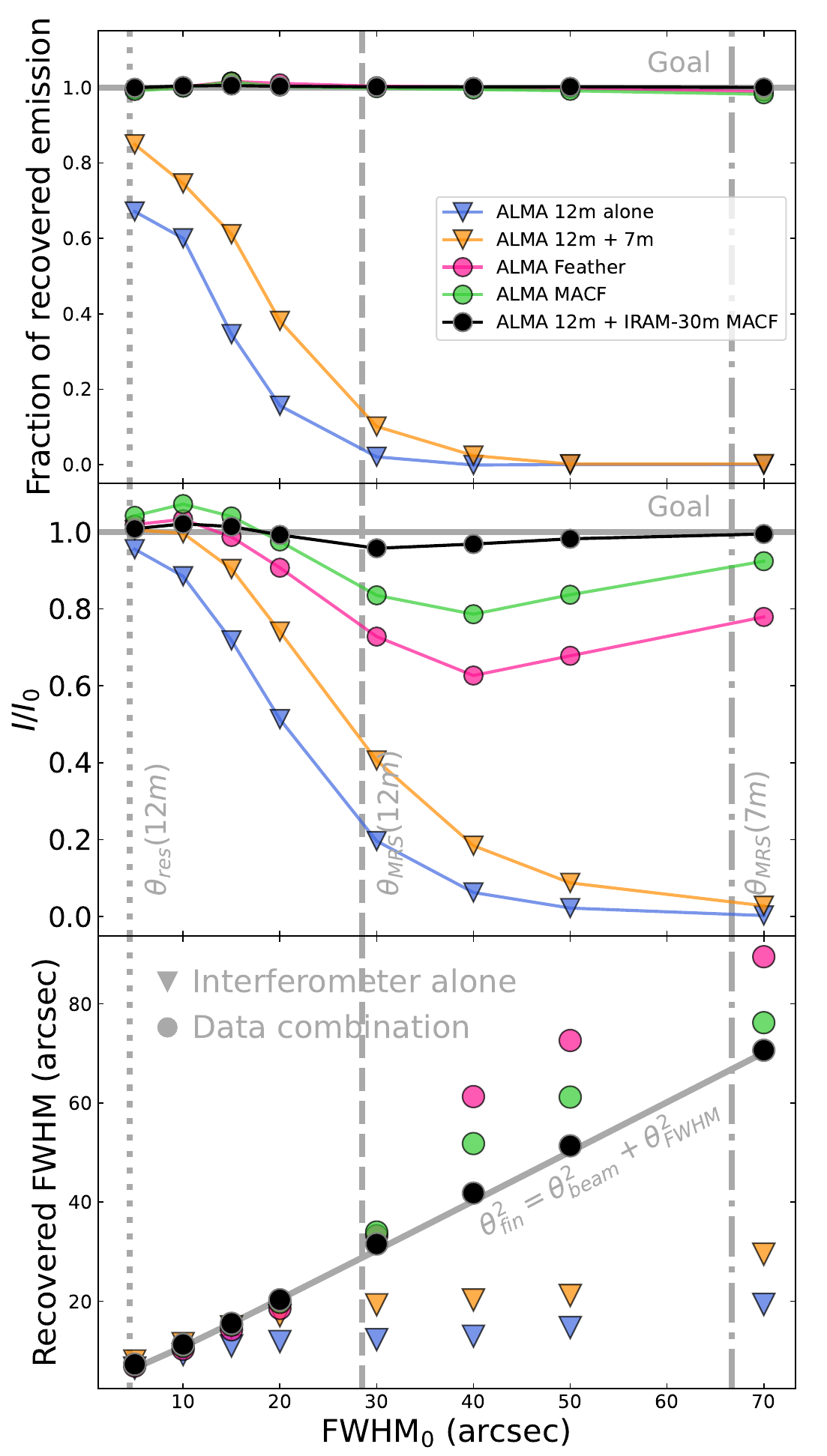}
	\caption{Reconstructed properties of isolated Gaussian filaments. \textbf{Top panel}. Percentage of emission recovered along the entire radial profile vs. FWHM$_0$ Different colors are used for the different data combination methods used: blue for the ALMA 12m alone, orange for the 12m + 7m, pink for the ALMA Feather, green for the ALMA MACF, and black for the ALMA 12m + IRAM-30m MACF. The grey dotted and dashed lines show $\theta_{res}(12m)$ and $\theta_{MRS}(12m)$, respectively. \textbf{Mid panel}. Intensity peak estimated from the Gaussian fit vs. FWHM$_0$. \textbf{Lower panel}. FWHM estimated from the Gaussian fit vs. FWHM$_0$. The grey solid line shows the theoretical expectation. For comparison, the results for the radial profile shown in Fig.~\ref{fig:gausfil_30arcsec_profile} with FWHM$_0$=30 arcsec correspond to the values close to the $\theta_{MRS}(12m)$}
	\label{fig:gauss_filaments_properties}
\end{figure}

To quantify the emission properties of our filaments we extract a radial profile from a cut along the x axis of each image, as done for cores. First, we estimated the total flux recovery integrating the total emission of each filament along their radial entire cut 
for filaments with different FWHM$_0$ values in Fig.~\ref{fig:gauss_filaments_properties} (\textit{top panel}).
Even for the narrowest filaments (FWHM$_0\sim$5~arcsec), the interferometer alone (blue solid line) is only able to recover $\sim70\%$ of the total emission of the source and, as the FWHM$_0$ grows, the fraction of recovered emission decreases also faster than in cores (see Fig.~\ref{fig:cores_properties}). Around FWHM$_0\sim \theta_{MRS}(12m)$ (grey dashed line), the ALMA 12m alone is losing almost 95$\%$ of the emission (note that the object is barely visible in Fig~\ref{fig:gausfil_comparison}). The ALMA 12m alone is clearly affected by the spatial filtering effect at all scales, not being able to recover the total emission even at smaller sizes. 
Adding intermediate-scale data (ALMA 12m + 7m; orange solid line) contributes to recover a larger fraction of emission at all cases. However the improvement is much smaller than for the cores, only recovering $\sim$20$\%$ of the emission around $\theta_{MRS}(12m)$. Even in this case it is clear how combining small and intermediate scales is not enough to fully recover these elongated structures.
Only the ALMA Feather and MACF (in pink and green lines, respectively) show constant profiles always recovering a fraction $\sim 100 \%$ of the total emission despite the filament size. These two profiles show the effects of combining interferometric data with SD observations, allowing to recover almost the total flux even at scales where the interferometric contribution is almost zero.

We continue fitting the extracted radial profile with a Gaussian function to estimate the intensity peak and the observed FWHM. The intensity peak normalized by the reference one $I/I_0$ is shown in  Fig.~\ref{fig:gauss_filaments_properties} (\textit{mid panel}). The behaviour of different data combinations is similar to what was observed for the recovered emission along the entire profile (see above). 
Both ALMA 12m alone (blue solid line) and ALMA 12m + 7m (orange solid line) data satisfactorily recover the peak emission (>95$\%$) in the case of unresolved filaments (FWHM=5 < $\theta_{res}(12m)$). On the other hand, both methods fail in the case of broader filaments as seen by their monotonically decreasing performance at larger FWHM$_0$. As expected, the filtering effects are more severe in the case of ALMA 12m-alone, although these issues are clearly visible in both datasets.
In comparison, the ALMA Feather (pink solid line) and ALMA MACF (green solid line) reductions show significant and consistent improvements at all scales. Still, the peak recovery at large FWHM values drops to $\sim$70-80$\%$ in both cases around 40~arcsec, growing again towards larger scales. Although secondary, we also notice that the ALMA MACF overestimates the peak intensity by few percent in the case of filaments with FWHM$_0\lesssim$15~arcsec. While the effects of the spatial filtering are still visible in all interferometric datasets, these issues are heavily reduced by the combination with the short-spacing information provided by the ALMA TP data\footnote{In all data combinations, we note that the peak recovery (\textit{mid panel}) is worse than the total flux recovery (upper panel) for filaments with similar FWHM. This non-intuitive result can be explained by the flux contribution of the sidelobes at large radii.}.

Finally, we display the FWHM recovered from our Gaussian fits compared to the inputs values in our simulations in Fig.~\ref{fig:gauss_filaments_properties} (\textit{lower panel}). 
Only the values recovered by the ALMA 12m alone (blue triangles) for a source FWHM$_0\le$15~arcsec lie on the expected prediction, while for larger objects the value of the recovered FWHM is systematically underestimated and becomes close to constant above FWHM$_0\sim$30~arcsec. At $\theta_{MRS}(12m)$ the estimated FWHM is around 10~arcsec, almost a factor of 3 lower than the reference value. 
The ALMA 12m + 7m points (orange dots) recover a slightly larger FWHM value closer to the theoretical prediction but where at FWHM$_0\sim \theta_{MRS}(12m)$ the FWHM is already underestimated by 30\%. 
The spatial filtering is thus affecting both the flux recovery and the object size. Filaments observed with the interferometer alone appear to be artificially narrower than expected.


In summary, filtering effects in elongated (filament-like) structures appear to be more severe than in the case of cores of similar FWHM$_0$ because the overall fraction of power at larger scales is much higher for filaments than for circular cores. These spatial contributions can significantly affect the observational (integrated intensity, peak intensity, and FWHM) and physical properties (i.e. total column density, peak column density, and radial distribution, respectively) derived in interferometric studies \citep[e.g., see Fig.1 in][]{Wong2022}.

\subsection{Finite filaments} \label{sec:el_cores}

\begin{figure}[htbp]
	\centering
	\includegraphics[clip=true,trim=0.0cm 0.0cm 0.0cm 0.0cm,width=\columnwidth]{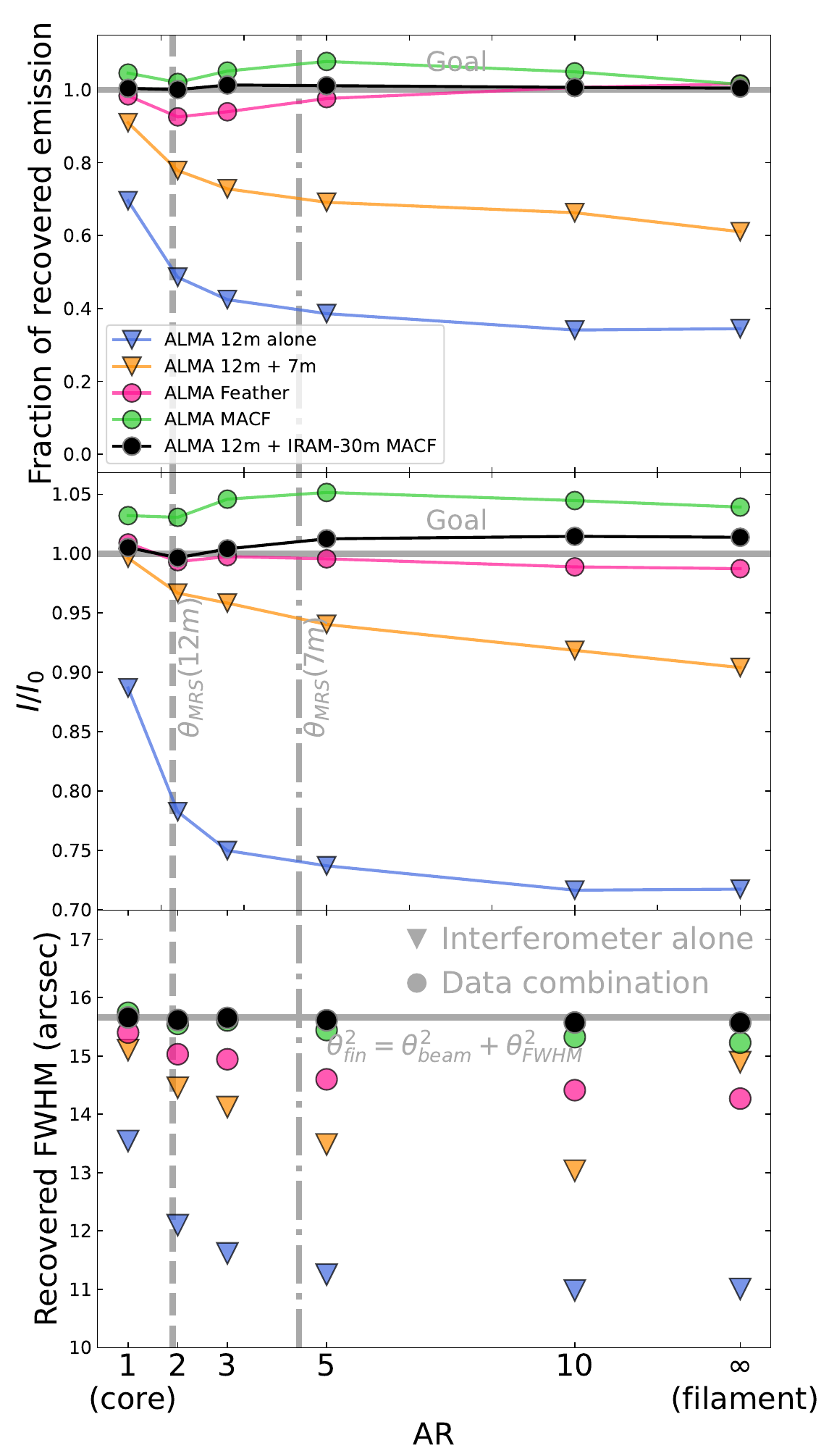}
	\caption{Reconstructed properties of noise-free isolated objects with FWHM$_0$=15~arcsec and aspect ratio AR=1 (core), 2, 5, and $\infty$ (infinite filament). From top to bottom: \textbf{(Top panel)} Fraction of recovered flux; \textbf{(Mid panel)} intensity peak estimated from the Gaussian fit; and \textbf{(Lower panel)} FWHM estimated from the Gaussian fit. 
    Different colors are used for the different data combination methods used: ALMA 12m alone (blue), 12m + 7m (orange), ALMA Feather (pink), ALMA MACF (green), and ALMA 12m + IRAM-30m MACF (black). The grey dotted and dashed lines show $\theta_{res}(12m)$ and $\theta_{MRS}(12m)$, respectively. 
    The grey solid line in the lower panel shows the theoretical expectation (FWHM$_0$=15~arcsec).}
	\label{fig:elcores_properties}
\end{figure}

Compared to the arbitrarily long, elongated structures simulated in Sect.~\ref{sec:fil} real filaments have finite lengths L. 
To investigate how this finite dimension is reproduced by interferometric observations, we decide to simulate a series of elongated objects with different aspect ratio ($AR=L/FWHM$).

We model these new targets as elongated 2D-Gaussians of different length described by Eq.~\ref{eq:core} where we independently vary $\sigma_x$ and $\sigma_y$ describing the characteristic FWHM and length L of the object, respectively.  For simplicity, we fixed the FWHM of our synthetic structures to 15~arcsec ($\sigma_x$) and vary only their length L ($\sigma_y$). Our choice for a constant FWHM=15~arcsec (similar to $\theta_{res}(12m)$) is justified as the characteristic width in which our previous core and filament models still recover most of the target properties but start diverging in their results (see Figs.~\ref{fig:cores_comparison} and \ref{fig:gausfil_comparison}). We produced ALMA simulations for different AR (or L), ranging from a core-like structure (AR=1) to an infinite-like filament ($AR=\infty$), using different arrays and combinations (see Fig.~\ref{fig:elcores_comparison}). 

Similar to our previous analysis, we extracted the main radial properties of our simulations from a cut along the x axis of each image and display these results in Fig.~\ref{fig:elcores_properties}. As expected, both the ALMA 12m alone (blue lines) and ALMA 12m+7m (orange lines) observations produce worse results than those datasets including TP information (i.e. Feather or MACF shown by pink and green lines, respectively) in terms of flux recovery (upper panel), peak intensity (mid panel), and recovered FWHM (lower panel). More interesting,  Fig.~\ref{fig:elcores_properties} smoothly connects the results obtained in the two limiting cases explored in previous sections.
Most variations in terms of flux, peak intensity, and FWHM already occur in mildly elongated core-like structures with $AR\sim 2-3$ while no further looses appear in more filamentary structures above $AR>5$. 

Although secondary with respect to the FWHM (see above), our results demonstrate how changes in the target's longest dimension L can aggravate the interferometric filtering even in structure with small $AR$. This is particularly relevant since dense cores usually show prolate geometries with $AR\sim 1.5-2$ \citep{1991Myers} inducing an additional source of uncertainty ($\sim20-40$\% extra losses) to previous observational biases (e.g., Sect.~\ref{sec:corenoise}).

\subsection{Filaments with Plummer-like profiles}

\begin{figure*}[htbp]
	\centering
	\includegraphics[clip=true,trim=0.cm 0.cm 0.cm 0.cm,width=\textwidth]{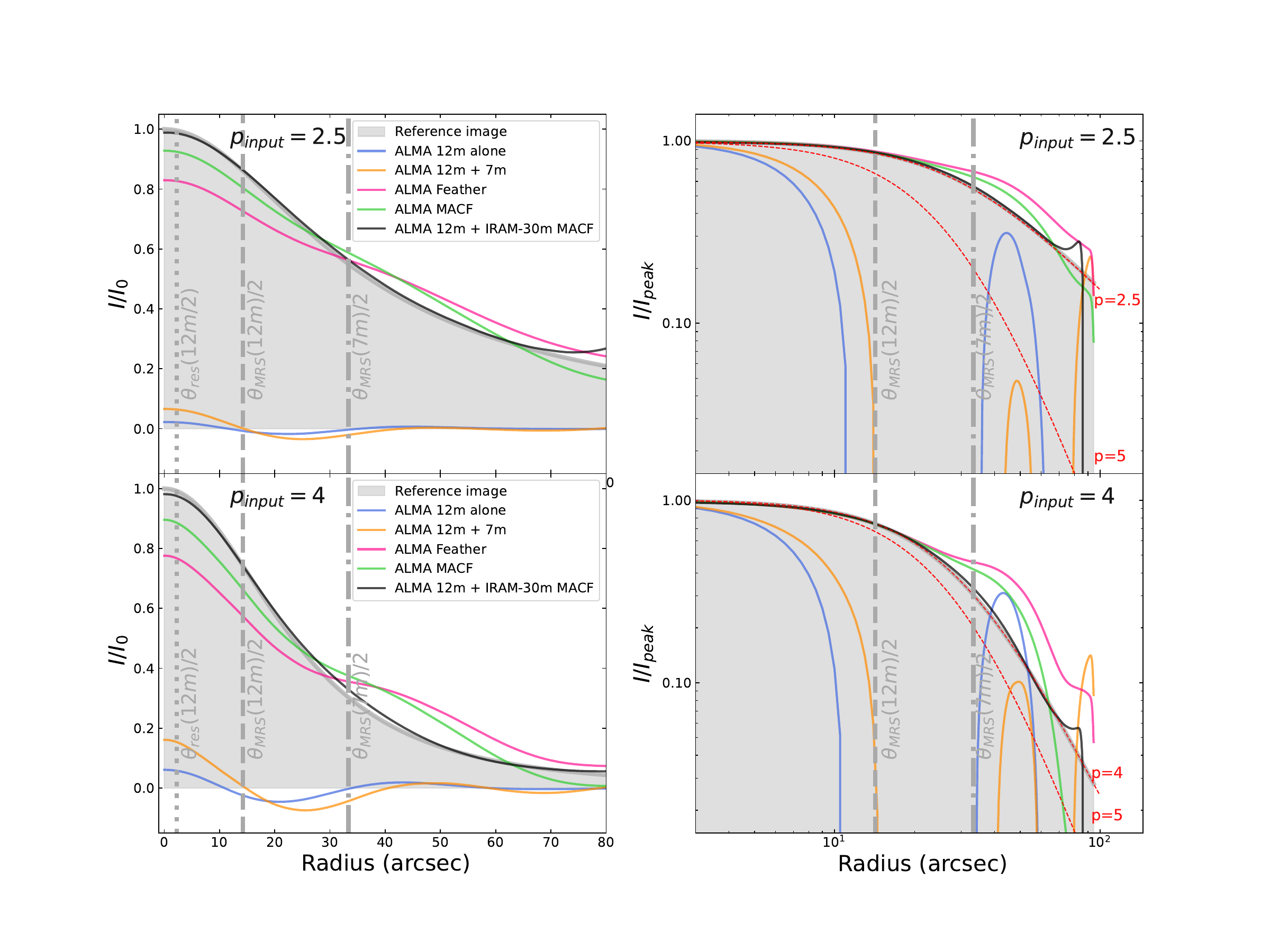}
	\caption{Radial profile extracted through a horizontal cut in the center of a p=2.5 (\textit{top panels}) and p=4 (\textit{lower panels}) $R_{flat}$=30~arcsec Plummer profile. \textbf{(Left panel)} Profile is normalized with respect to the reference peak and displayed in normal scale. \textbf{(Right panel)} Profile normalized with respect to its peak value displayed in log-scale. The data combination methods use are marked in different colors: ALMA 12m alone (blue), ALMA 12m + 7m (orange), ALMA Feather (pink), ALMA MACF (green), and ALMA 12m + IRAM-30m MACF (black). The reference profile is marked by the gray shadowed area and ideal profiles for different p-values are displayed by red dashed lines.}
	\label{fig:plummer_filaments}
\end{figure*}

Since filaments observed with \herschel~show a characteristic Plummer radial profile \citep{2011Arzoumanian}, we decided to reproduce these more realistic radial profiles in our analysis of filaments. We modelled the radial profile as constant along the y-axis and a Plummer-like profile along the x-axis, described as
\begin{equation}
    N(H_2)|_p(r)=A_p \frac{\rho_c\ R_{flat}}{[1+(r/R_{flat})^2]^{\frac{p-1}{2}}}
\end{equation}
where $A_p=\frac{1}{\cos{i}}\int_{-\inf}^{+\inf} \frac{du}{(1+u^2)^{p/2}}$, $\rho_c$ is the density at the center, $R_{flat}$ is the characteristic radius of the flat inner portion of the profile, and $i$ inclination angle assumed to be equal to 0 for simplicity \citep[see][]{2011Arzoumanian}. 

After investigating how the spatial filtering affects the flux recovery and the FWHM estimate in the case of Gaussian filaments, in Sect.~\ref{sec:gaussianfils}-\ref{sec:el_cores}, here we decided to focus on the effects on the profile slope. 
Thus, we produce two noise-free, Plummer-like filaments with p=2.5 and 4 as representative values for shallow and steep filaments, respectively \citep[see][]{2024Hacar}, both with radii of $R_{flat}$=30~arcsec and constant peak flux value equivalent to I$_0$= A$_p\times\rho \times R_{flat} =7.5\times10^{-3}$ mJy/pix.
Compared to Sect.~\ref{sec:fil}, the selected p values describe the power-law dependence of a steep, Ostriker-like filament, similar to a Gaussian (p=4; \citet{1964Ostriker}), and the much shallower Plummer-like variations similar to those reported by Herschel \citep[p=2.5, see][]{2011Arzoumanian}.
Accurately recovering these filament profiles, which differences are only noticeable at large radii (r$\gg$FWHM), is essential to determine the evolutionary state of these structures \citep{Pineda2023}.

In Fig.~\ref{fig:plummer_filaments}, we display the radial profiles extracted from a horizontal cut in the two filaments with p=2.5 (\textit{upper panels}) and p=4 (\textit{lower panels}). We show both the normalized intensity I/I$_0$ (in linear scale; \textit{left panels}). The observation of these Plummer-like profiles retrieves worse results than Gaussian-like filaments of similar angular size (Sect.~\ref{sec:gaussianfils}). These new shallower profiles enhance the effective filtering reducing the observed peak intensity in our ALMA 12m alone (blue line) and ALMA 12m + 7m profile (orange) simulations, particularly in the case of p=2.5 (recovering only 5$\%$ of the original peak). In addition to their poor performance, we note that interferometric alone observations may also introduce significant sidelobes with intensities comparable to the main filament peak (see secondary peaks in ALMA 12m alone and ALMA 12m + 7m datasets. On the other hand, only combination methods such as ALMA Feather (pink line) and ALMA MACF (green line) get closer to the true emission profiles of these filaments, although not in a complete satisfactory way (see Sect.~\ref{sec:psf} for a further discussion).

More important, and as highlighted in the normalized I/I$_{peak}$ plots in Fig.~\ref{fig:plummer_filaments} (\textit{right panels}), the previously reported flux looses have a large impact on the resulting radial profiles measured in these filaments. In order to better evaluate the profile's slope, different dashed red lines show the expected Plummer dependence for the respective p=2.5 or 4 values (solid red line) and a representative steeper profile with p=5 (dashed red line). The recovered ALMA 12m alone profiles are much sharper than the original ones showing power-law dependencies significantly steeper than p=5. 
In addition to their large flux differences, filtering severely impacts the resulting radial profiles producing artificially sharp filaments in all interferometric-alone observations (both ALMA 12m alone and ALMA 12m + 7m). Recent results illustrate the impact of these interferometric filtering effects in ALMA observations of different filamentary structures in the ISM \citep{Klaassen2020,2021Yamagishi,2023DiazGonzalez,2024Hacar,2024Tachihara}.
These observational biases should be considered in future ALMA studies aiming to characterize the radial dependence of the ISM filaments at high spatial resolution \citep[e.g., ][]{2024Socci}.

\section{Point-Spread-Function (PSF) analysis} \label{sec:psf}

In addition to the source geometry, the recovery of the true sky emission distribution is limited by the corresponding Point Spread Function (PSF) in interferometric observations.
The interferometer PSF is the inverse Fourier transform of the u-v components sampled during the observation. The number of antennas, its configuration, and the integration time during an observation determine the u-v coverage and, therefore, the resulting PSF. Considering an ideal case with an uniformly sampled and homogeneous u-v coverage, the PSF would appear as a 2D Gaussian with a FWHM value defining the interferometer resolution. However, since the u-v sampling is always discrete and non homogeneous, realistic PSF can largely depart from the idealized Gaussian functions. Combined with the missing short-spacing information (central hole in the uv-plane), observations carried out with a limited u-v coverage can lead into a PSF with prominent sidelobes that can severely hamper the deconvolution process.

\begin{figure}[ht!]
	\centering
	\includegraphics[clip=true,trim=0.0cm 0.0cm 0.0cm 0.0cm,width=1\linewidth]{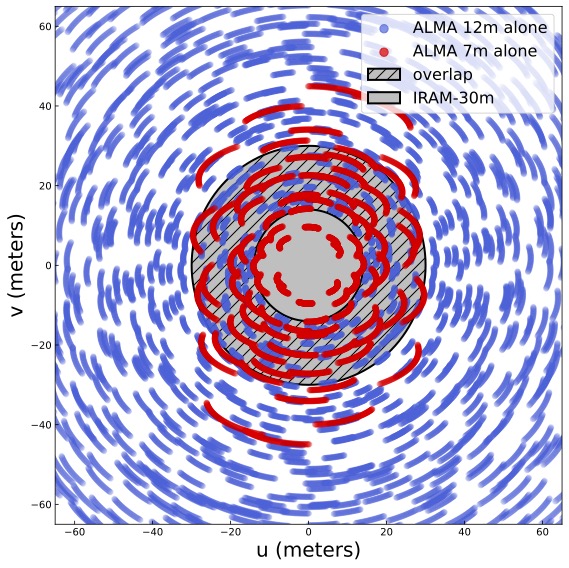}
	\caption{uv-coverage of one of the cores simulated in Sect.~\ref{sec:method}. Different colors represent the uv-sampling obtained by the different ALMA arrays: (blue) ALMA 12m, (red) ACA 7m array, and (orange) TP.
		We also highlight the uv-area covered by the IRAM-30m SD telescope (grey area) and its overlap with the ALMA 12m array (black dashed area).}
	\label{fig:uv_coverage}
\end{figure}

In Fig.~\ref{fig:uv_coverage} we show the u-v components measured during the observations of one of our Gaussian cores (see Sect.~\ref{sec:cores}). Since we adopted the same observational setup for all simulations (see Sect.~\ref{sec:method}), the u-v coverage and the resulting PSF are similar in all the cases presented so far. Thanks to large number of antennas, the ALMA 12m array (blue) presents a dense instantaneous (snapshot) coverage that populates the u-v space regularly and homogeneously down to baselines of $\sim$14m (i.e. the shortest baseline in this setup, see Sect.~\ref{sec:method}). In contrast, the more limited number of ACA 7m antennas (red) create more sparse u-v components usually distributed in a highly non symmetric pattern (elongated toward the v-direction in our simulations) also leaving large u-v holes between points (e.g., see (u,v)$\sim$(20,30) meters).
Given their different observational setups, we therefore expect strong departures between the PSF of the 12m and 7m arrays.

\begin{figure*}[htbp!]
	\centering
	\includegraphics[clip=true,trim=0.0cm 0.0cm 0.0cm 0.0cm,width=0.97\textwidth]{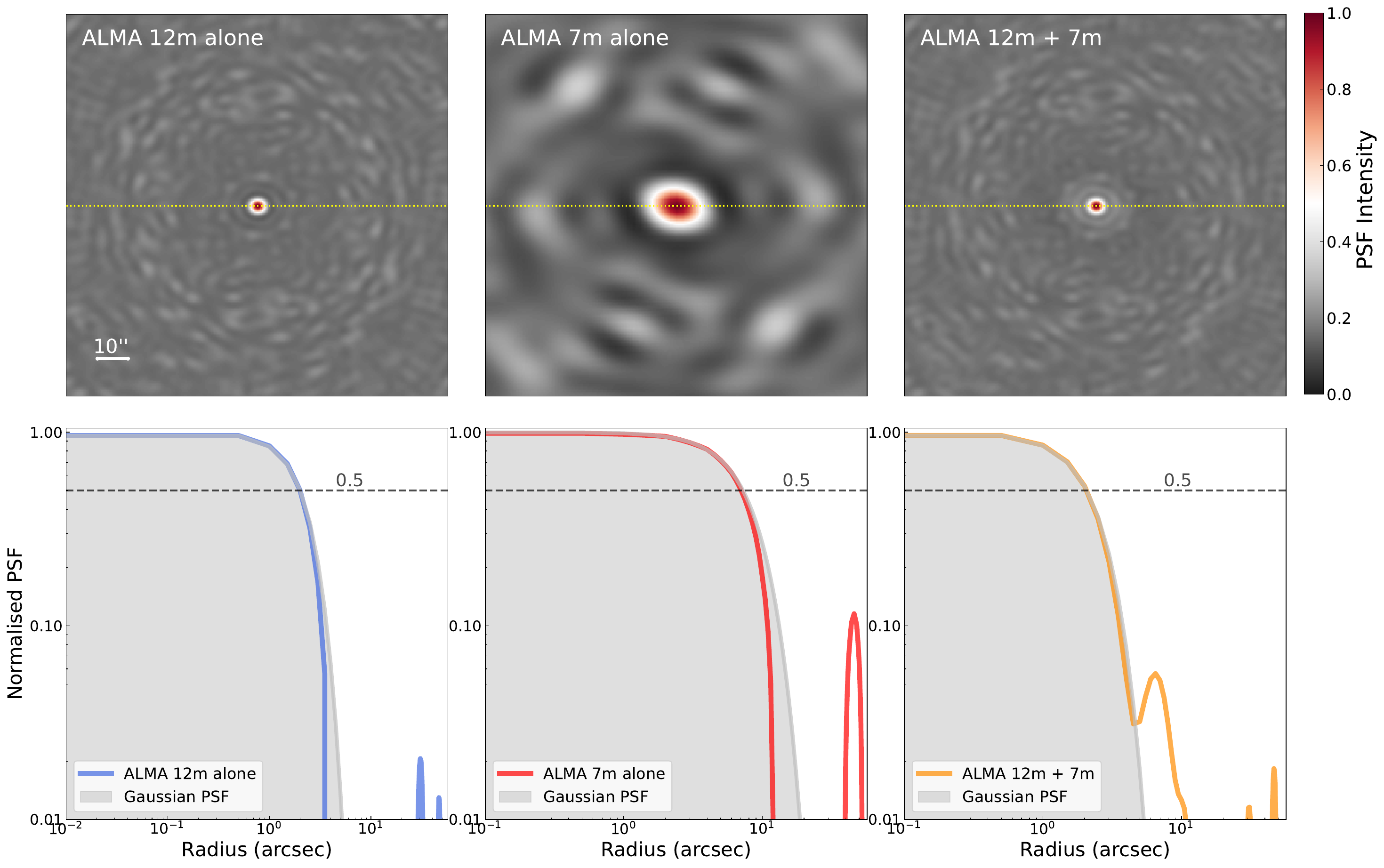}
	\caption{Analysis of the interferometer PSF of different ALMA array configurations. From left to right:
		 \textbf{(left panels)} ALMA 12m alone ,  \textbf{(middle panels)} 7m alone, and \textbf{(right panels)} ALMA 12m + 7m. 
		 We display the PSF intensity image for each simulation \textbf{(top)} and an intensity profile extracted along a horizontal cut in log-scale \textbf{(bottom)}. In the bottom panels, the colored lines show the simulated PSF profiles (see dotted yellow lines in mid-panels) obtained for the ALMA 12m alone (blue), 7m alone (red) and 12m + 7m (orange), while the expected Gaussian PSF profile at the correspondent resolution is marked by the shaded grey area.}
	\label{fig:psf}
\end{figure*}

In Fig.~\ref{fig:psf} we show the image (\textit{top row}) and a representative radial cut of the interferometric PSF (\textit{bottom row}) obtained in our ALMA 12m alone (\textit{left panels}), 7m alone (middle panels) and 12m + 7m (\textit{right panels}) simulations.
The ALMA 12m alone PSF (top left panel) shows a quasi-Gaussian shape ($b_{maj}$=4~arcsec, $b_{min}$=3.5~arcsec) surrounded by a concentric, regular and symmetric sidelobe pattern. 
In comparison, the ALMA 7m alone PSF (top middle panel in Fig.~\ref{fig:psf}) presents a more elongated quasi-Gaussian shape ($b_{maj}$=14.5~arcsec, $b_{min}$=10.3~arcsec) surrounded by an irregular and asymmetric pattern of brighter sidelobes.
We extracted a radial profile at the center of the ALMA 12m PSF image. We plotted it in log-scale (bottom panels) and compared them with their corresponding ideal Gaussian PSF such as $\theta=b_{maj}$ (gray dashed area). 
As expected, the ALMA 7m alone PSF is significantly worse than the ALMA 12m alone one, showing an earlier departure from gaussianity as well as brighter sidelobes ($\sim15\%$ vs $\leq2\%$ of the PSF peak).

\begin{figure}[htbp]
	\centering
	\includegraphics[clip=true,trim=0.0cm 0.0cm 0.0cm 0.0cm,width=1.0\linewidth]{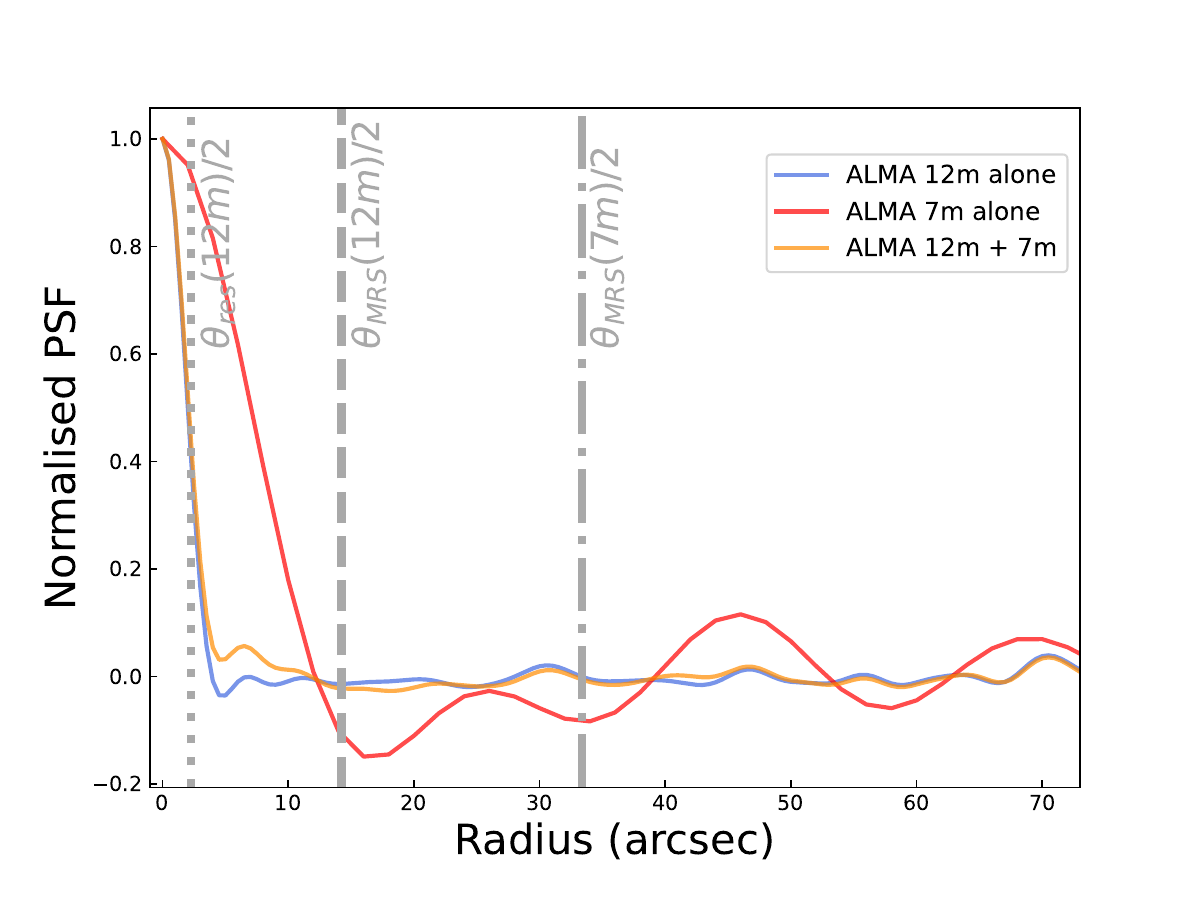}
	\caption{Comparison of the interferometer PSF intensity profile extracted from an horizontal cut for the ALMA 12m alone (blue), 7m alone (red) and 12m + 7m (orange). The resolution of the 12m array (dotted line), and the MRS of both the 12m (dashed line) and 7m (dot-dashed line) arrays are indicated by vertical lines.}
	\label{fig:psf_comparison}
\end{figure}

Naively, we would expect the combination of ALMA 12m + 7m baselines to improve the overall PSF shape. As shown in the right panels in Fig.~\ref{fig:psf}, the 12m+7m combined PSF shows a well-behaved Gaussian-like dependence at small radii with a reduced influence of the sidelobes seen in the 7m-alone data. 
However, the addition of 7m data also introduces a strong and unexpected but prominent sidelobe at scales around 6~arcsec, and around the overlapping baselines shown in Fig.~\ref{fig:uv_coverage}, with an estimated peak intensity around 7$\%$.

The observed ALMA 12m + 7m PSF can lead into the redistribution of the emission during the deconvolution process of bright extended sources by placing additional flux on the sidelobes. This effect might affect the results of our ALMA Feather and ALMA MACF simulations, and likely explains their discrepancies observed in the flux peak and FWHM recovery with respect to the expected sky emission (also around 10-20$\%$, see Fig.~\ref{fig:cores_properties} and \ref{fig:gauss_filaments_properties}). 

Deconvolution algorithms such as CLEAN efficiently mitigate most of the above PSF issues in the case of simple source geometries. These improvements are however more limited when targeting fields with complex emission features such as those explored in our EMERGE survey (see Paper I). Depending on the source structure, relatively minor sidelobes ($\sim$10\%) could lead into complex dirty images from which CLEAN can no longer recover. 
While a full analysis of these PSF effects is outside the scope of this paper, optimizing the PSF quality appears crucial for achieving interferometric images with accurate flux measurements better than 20\% in ISM studies.

\section{Data combination with a large SD: ALMA + IRAM-30m observations}\label{sec:iram}

The analysis presented in this work demonstrates how the inclusion of the short-spacing information using data combination systematically improves the quality of the data allowing us to retrieve the true properties of the both cores (Sect.~\ref{sec:cores}) and filaments (Sect.~\ref{sec:fil}) observed with ALMA. However, the results of the ALMA Feather and MACF methods combining the three ALMA arrays are still not ideal. Indeed, differences in the recovered flux peak (underestimated) and FWHM (overestimated) are clearly visible in Fig.~\ref{fig:plummer_filaments} (among others), this discrepancy can be partially attributed to the resulting PSF (see Sect.~\ref{sec:psf}). 
Thus, we wanted to explore whether the ALMA 12m alone would benefit more from the combination with a larger SD compared to the ALMA one. Such improvements are expected by the more homogeneous uv-coverage obtained by a large SD with respect to a small interferometer such as the ALMA (ACA) 7m array \citep[see][]{2023Plunkett}. 

We simulate the observations of a SD telescope larger than the ALMA TP (12m antennas) to be combined with the ALMA 12m array, replacing the ACA-7m and the TP. 
We chose to reproduce the 30-meter Institute de Radioastronomie Millimetric telescope (IRAM-30m), commonly used in star-formation studies also included as part of our EMERGE survey (see also Sect.~\ref{sec:real data}). In order to simulate an IRAM-30m observation, we convolve the synthetic image at the resolution of 25~arcsec ($\sim$ resolution of IRAM-30m telescope at 100 GHz) using the CASA task \texttt{imsmooth} and combined it with ALMA 12m array observations using the MACF method (see Sec.~\ref{sec:method}), hereafter referred to as ALMA 12m + IRAM-30m observations.

In agreement to our predictions, the ALMA 12m + IRAM-30m MACF appears as the best data combination method. A quantitative analysis of their results indicate that the ALMA 12m + IRAM-30m MACF observations accurately reproduce the total flux, peak intensities, and FWHM of both Gaussian cores (Fig.~\ref{fig:cores_properties}) and filaments (Figs.~\ref{fig:gauss_filaments_properties} and \ref{fig:plummer_filaments}) better than any previous combination within the ALMA arrays (ALMA Feather or ALMA MACF) and within less than $\sim$10\% with respect to the actual model values. 
Considering Plummer-like filaments (Fig.~\ref{fig:plummer_filaments}), arguably the most challenging targets for interferometers given their shallow profiles (p=2), the ALMA + IRAM-30m MACF profile is the only method closely reproducing the synthetic filament. Even if the ALMA Feather and MACF reproduce the true slope in profiles with both p=2.5 and p=4 up to $\sim30$~arcsec as accurately as the ALMA + IRAM-30m MACF, the latter is still the only one recovering the intensity peak at the center.

The reason for these improvements is two fold. First, a large SD provides a high-sensitivity within a larger range of scales. Second, and more relevant for ALMA, the use of a SD provides a uniform uv-coverage of the short-spacing baselines that, unlike the ACA, does not introduce additional sidelobes that perturb the analysis (see Sect.~\ref{sec:psf}).

Overall, the analysis performed in this work suggests the combination of ALMA 12m with a larger SD (like the IRAM-30m) allow to achieve better results compared to using ALMA ACA-7m array + TP, always recovering the true properties of the sources despite of their size and shape. The combination with a large SD appears essential to achieve images with high fidelity even with densely populated interferometers such as ALMA \citep[see also][]{2024Hacar}. An enhanced performance is expected for future (50m-class) SD telescopes such as the Atacama Large Aperture Submillimetre Telescope (AtLAST) telescope \citep{2019Klaassen}, in particular in the case of ISM studies \citep{Klaassen2024}.

\subsection*{On the SD sensitivity for data combination}\label{sec:noise}

\begin{figure*}[htbp!]
	\centering
	\includegraphics[clip=true,trim=0.0cm 0.0cm 0.0cm 0.0cm,width=\textwidth]{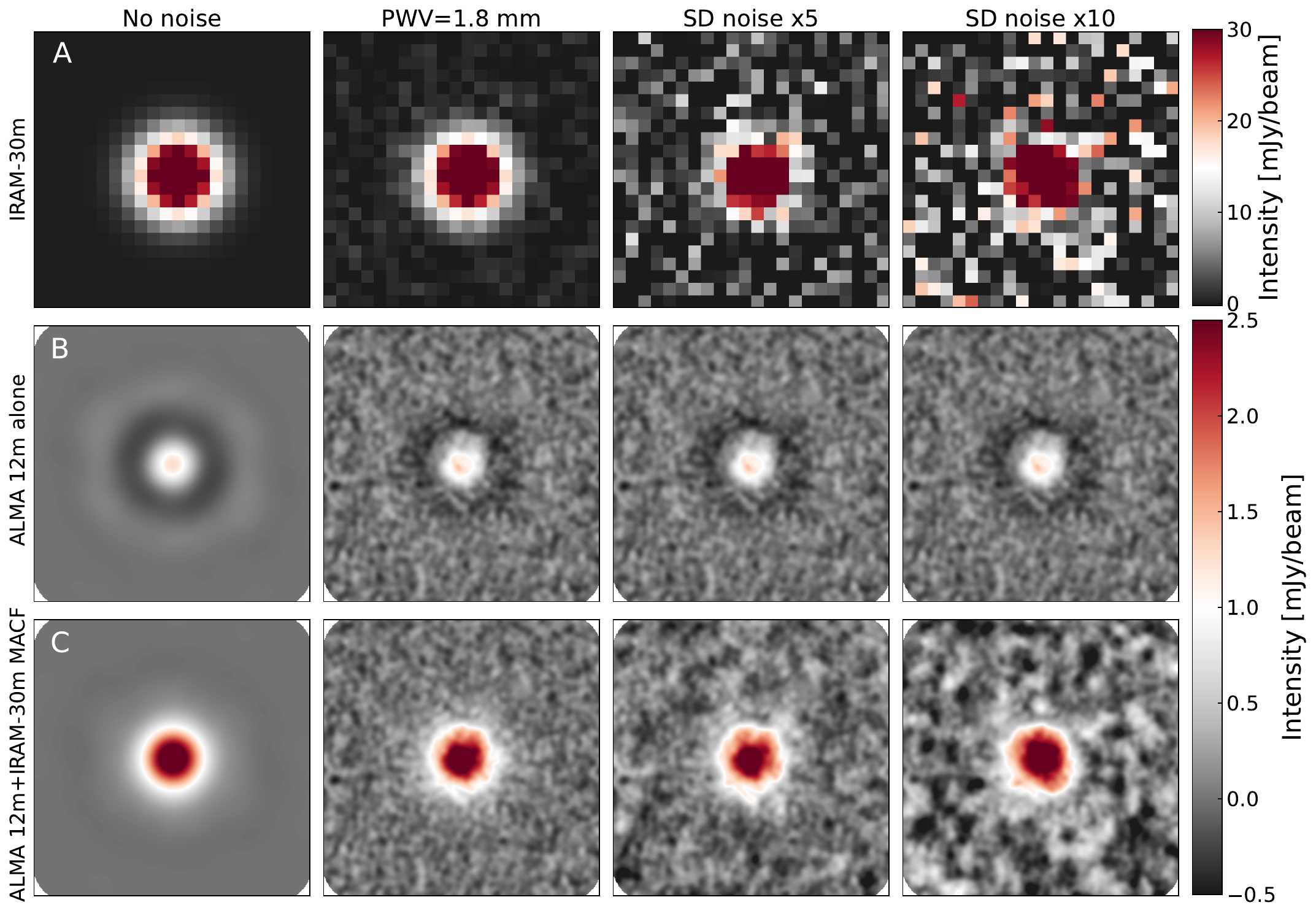}
	\caption{\textbf{From left to right:} Simulations of isolated cores with FWHM$_0$= 30~arcsec. From left to right: noise-free simulations, and simulations with noise levels equal to 1, 5, and 10 times the theoretical noise (assuming PWV=1.8~mm). From top to bottom: IRAM-30m simulations (panel A), ALMA 12m alone (panel B), ALMA 12m + IRAM-30m MACF (panel C).}
	\label{fig:noise}
\end{figure*}

\begin{figure}[htbp!]
	\centering
	\includegraphics[clip=true,trim=0.0cm 0.0cm 0.0cm 0.0cm,width=\linewidth]{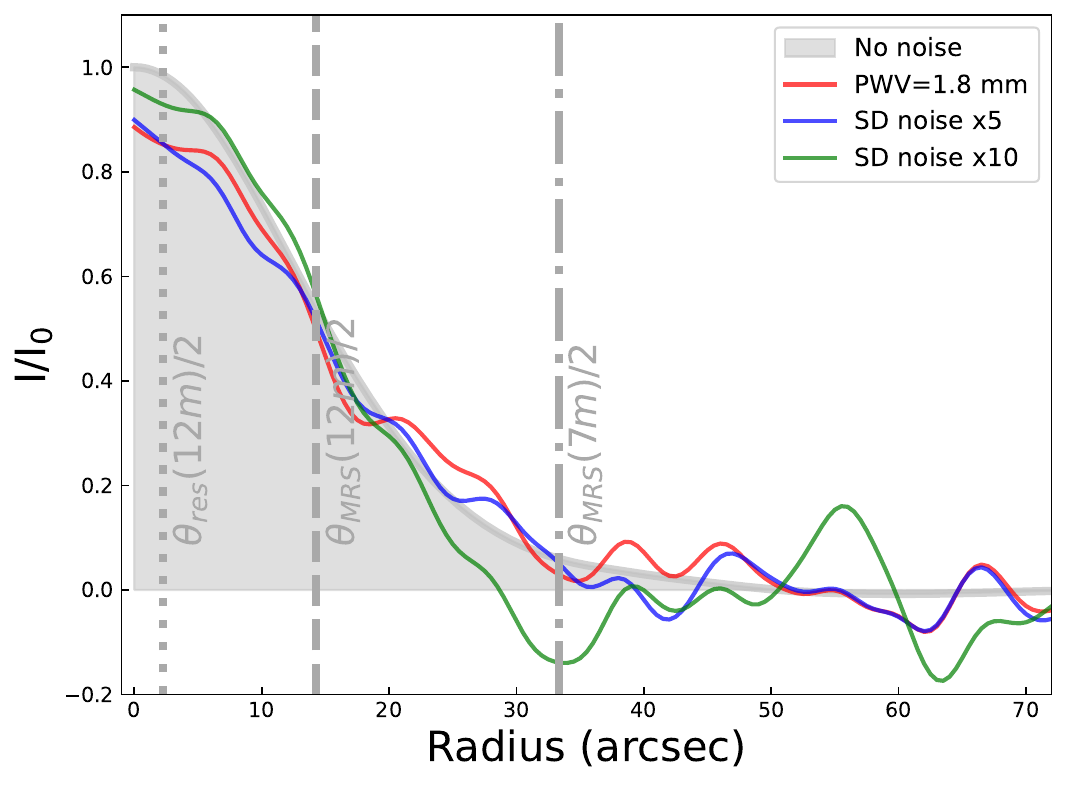}
	\caption{Radial profile extracted from an horizontal cut in the center of a FWHM$_0$=30~arcsec Gaussian cores shown in Fig.~\ref{fig:noise}, panel C.}
	\label{fig:noise_profile}
\end{figure}

When combining data between interferometric arrays and SD, a standard of good practice is to match the sensitivity for SD data to the one of the interferometer within the range of overlapping baselines \citep[see][]{2013Mason}. This requirement translates in longer integration times per pointing for the SD compared to the interferometric observations. Given the large telescopes times required \citep[e.g., 1:11.9 for the 12m:TP data in C43-1][]{2023Cortes}, it is interesting to test whether this theoretical requirement can be actually relaxed. 

In order to pursue this goal, we make use of our synthetic datasets including thermal noise (Sect.~\ref{sec:method}). 
Following \citet{2013Mason}, we determined the expected ratio of integration time per pointing ($\tau$) necessary to match the sensitivity between the ALMA 12m array (int) and the IRAM-30m telescope (SD) as follows:\footnote{These estimates assumed to observe with the same spectral setup, antenna efficiencies, and system temperature.} 
\begin{equation}
    \frac{\tau_{SD}}{\tau_{int}}=\left( \frac{D_{int}}{D_{SD}}\right)^4\ \frac{2N_{bas}}{N_{SD}}
\end{equation}
with D the diameters of the antennas, N$_{SD}$ number of SD antennas and N$_{bas}$ number of overlapping baselines (i.e.,the number of baselines falling within the SD diameter). Considering the ALMA 12m array in configuration C43-1 in Cycle 9 and IRAM-30m, $\tau_{SD}\sim4\tau_{int}$. Given the integration time ratio, we calculated the equivalent (white) noise in our IRAM-30m simulations following the radiometer equation. We adopt a standard Gaussian core with $FWHM_0=$~30~arcsec as illustrative example were data combination becomes essential (i.e. $\sim$~60\% of flux losses in the 12m alone observations; Sect.~\ref{sec:cores}).
We combined the ALMA 12m alone data and the IRAM-30m data using the MACF method.
We simulate these observations setting the SD noise as 1, 5, and 10 times the above theoretical predictions.

Different panels in Fig.~\ref{fig:noise} display the results of the IRAM-30m (panel A), ALMA 12m alone (panel B), and ALMA 12m + IRAM-30m MACF combination (panel C), respectively. Two features become apparent in these images. First, the increase of the SD noise rapidly degrades the IRAM-30m images (top panels) which also translates in the systematic increase of the noise in the combined MACF images (bottom panels). Second, and more important, a moderate increase of the SD noise level (up to 5 times) does not affect the recovery of the extended emission unless the SD image is heavily corrupted (e.g., 10 times the expected noise).
We illustrate these ALMA 12m + IRAM-30m MACF source profiles shown in Fig.~\ref{fig:noise_profile}. Even when the noise significantly increases, the recovered characteristics of the target source (i.e. flux peak, FWHM, and radial profile) remain stable (within 10-20\%). 

The above results could be explained by different contributions of the interferometric and SD data during combination.
Our ALMA interferometric observations determine the sensitivity at most spatial scales given the amount of ALMA 12m baselines (including several overlapping with the scales traced by the SD data), and drive the final noise level in the combined maps.
On the other hand, SD data mostly provide information of the low frequency (spatial) components. These (Fourier) components are still recognizable in images with high frequency noise levels (Fig.~\ref{fig:noise}, panel A) and therefore can still efficiently contribute to the recovery of extended emission (Fig.~\ref{fig:noise}, panel C)\footnote{We note that these conclusions might hold only in the case of SD data including random, high frequency noise. The addition of correlated noise with additional low frequency components (e.g., atmospheric variations during the SD observations) might corrupt the combined images even at low noise levels.}.

Our analysis might have interesting consequences for future ALMA observations. Since the image noise $\sigma_{RMS} \propto \tau^{-1/2} $, the SD observing times could be reduced without significantly compromising the final outcome of data combination. Also, the addition of short-spacing information appear to have a positive effect on the final data products, even if the original SD data are of lower quality than expected.

\section{Comparisons with real ALMA observations}\label{sec:real data}

\begin{figure*}[htbp]
	\centering
	\includegraphics[clip=true,trim=0.0cm 0.0cm 0.0cm 0.0cm,width=\textwidth]{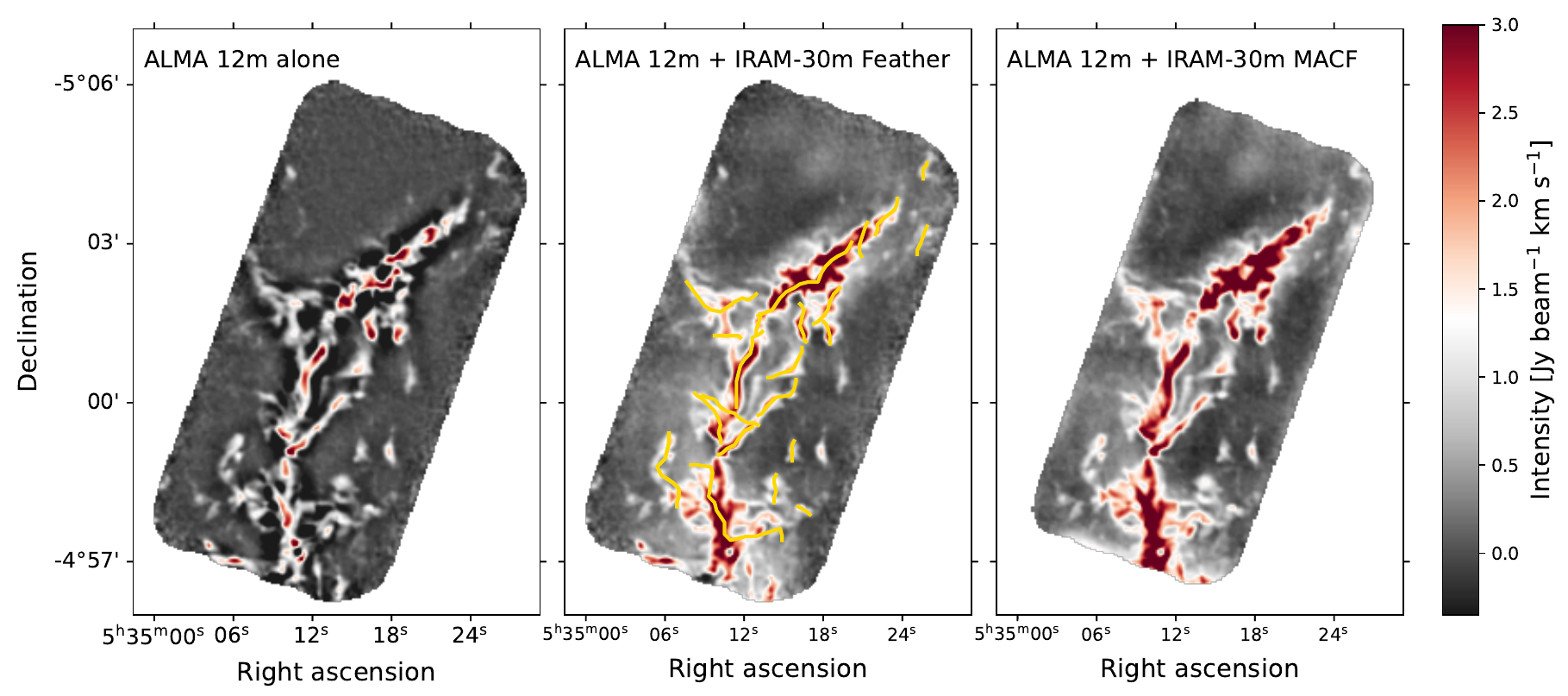}
	\caption{\nthp~integrated intensity maps of OMC-3 using three different data combination methods. From left to right: {\bf (Left panel)} ALMA 12m alone map; {\bf (Mid panel)}  ALMA 12m array + IRAM-30m combined using the Feather with the identified filament axes over-plotted (yellow; see also Paper III); and {\bf (Right panel)}  ALMA 12m array + IRAM-30m combined using the MACF method.}
	\label{fig:OMC-3_map}
\end{figure*}

We compare our synthetic simulations with real ALMA observations to validate our results. As target for these comparisons we selected the OMC-3 star-forming region in Orion A, part of the EMERGE Early ALMA Survey (see Paper I). OMC-3 shows a prominent and complex substructure of filaments and cores becoming an ideal testbed for our simulations. The dense gas content of OMC-3 has been investigated using \nthp (Band 3) observations carried out with the ALMA 12m-array (alone) in its most compact configuration (C43-1) during ALMA Cycle 7, achieving a native resolution of $\theta_{beam}\sim$~3.5~arcsec (proj. ID: 2019.1.00641.S; PI: Hacar).
The ALMA 12m array observations have been combined with additional IRAM-30m \nthp~large-scale maps (proj. IDs: 032-13, 120-20), using both the Feather and MACF methods and are referred to as ALMA 12m + IRAM-30m Feather and MACF, respectively. The observations, the data reduction and imaging procedures, and the data combination technique used are presented in \citet{2024Hacar}. Interestingly, the noise level of our IRAM-30m observations \citep{2017Hacarb} are higher than those expected according to theoretical expectations (see Sect.~\ref{sec:iram}).

We show the integrated intensity maps of all interferometric alone (ALMA 12m alone) and combined datasets (ALMA 12m + IRAM-30m Feather and MACF) in different panels in Fig.~\ref{fig:OMC-3_map}. The comparison between these maps clearly demonstrates how the ALMA 12m alone one (\textit{left panel} in Fig.~\ref{fig:OMC-3_map}) is not able to recover the diffuse emission, showing narrower and fainter filamentary structures surrounded by negative emission in agreement with the predictions of our simulations.

\begin{figure}[ht!]
	\centering
	\includegraphics[clip=true,trim=0.0cm 0.0cm 0.0cm 0.0cm,width=\columnwidth]{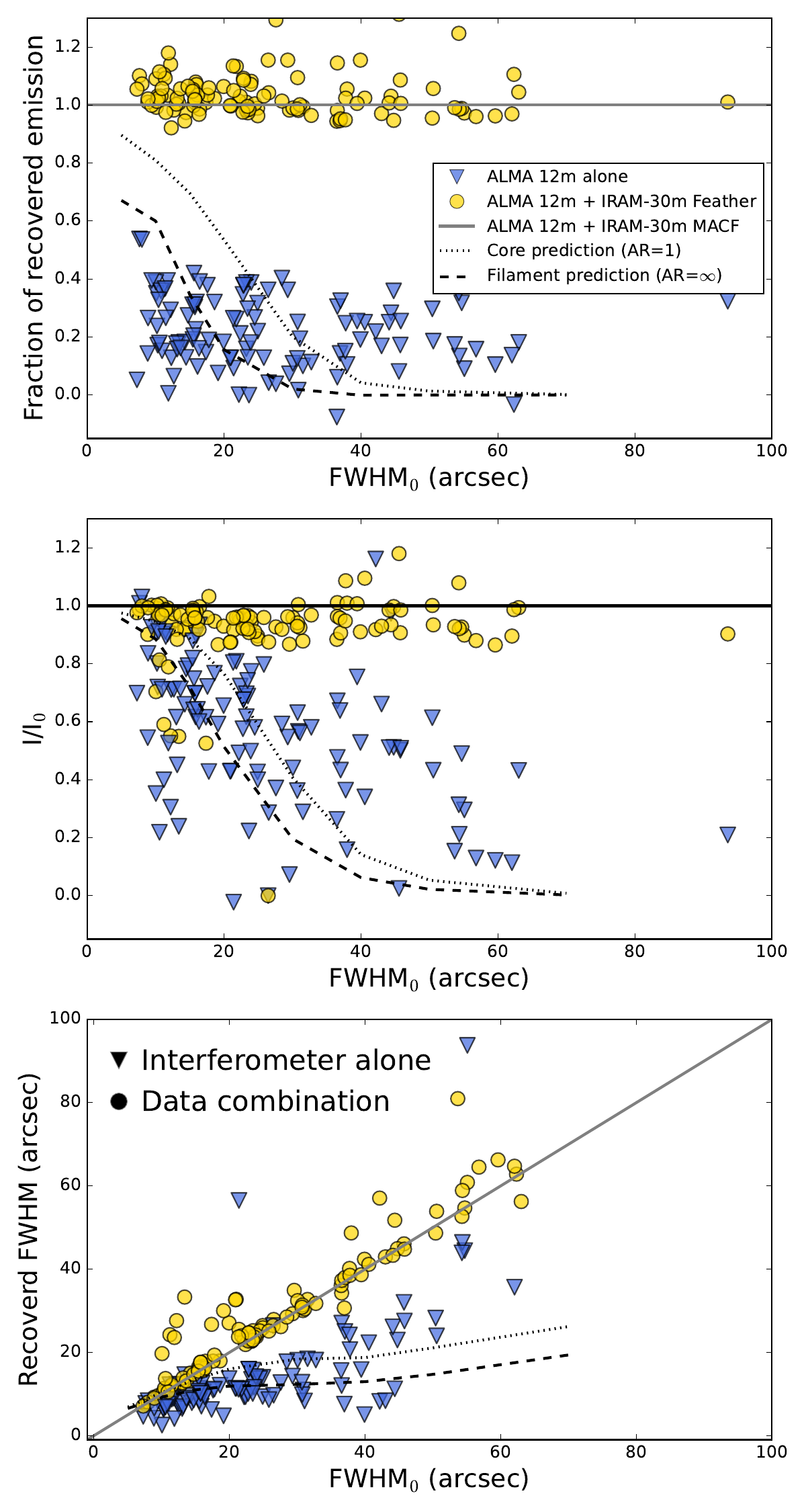}
	\caption{Properties of real filaments (total flux, peak flux, and FWHM) identified in the \nthp~integrated intensity maps of OMC-3 (see Fig.~\ref{fig:OMC-3_map}). All panels display the values obtained from the
		ALMA 12m alone (yellow dots) and ALMA 12m+ IRAM-30m Feather (blue triangles) 
		compared to those obtained form the combination of the ALMA 12m array + IRAM-30m  using the MACF (flux, I$_0$, and FWHM$_0$) method assumed as reference. A perfect agreement between these methods is indicated by a grey line.
		Predictions form simulations are displayed by the grey dashed line. 
		From top to bottom:
		\textbf{(Top panel)} Percentage of emission recovered along the entire radial profile. 
		\textbf{(Mid panel)} Intensity peak FWHM estimated from the Gaussian fit. 
		\textbf{(Lower panel)} FWHM estimated from the Gaussian fit.
		In all cases, the values are represented as function of the reference filament FWHM$_0$ derived from the ALMA 12m array + IRAM-30m MACF map.
	}
	\label{fig:OMC-3_analysis}
\end{figure}

We have characterized the emission in these maps from the analysis of the 23 small-scale filaments (aka fibers; marked by segments on the Fig.~\ref{fig:OMC-3_map}, \textit{mid panel}) identified in this region (see Paper III for additional details). In order to compare the results on these data with the simulated ones, we extract several radial cuts perpendicular to each fiber axis and analyse their radial profiles. Fig.~\ref{fig:OMC-3_analysis} shows a statistical comparison of the flux (\textit{top panel}), peak flux (I$_0$, \textit{mid panel}), and FWHM (\textit{lower panel}) recovered in these images. 
Without a prior knowledge of the true emission distribution, and given the results in Sect.~\ref{sec:iram}, we assume the ALMA 12m + IRAM-30m MACF observations as reference and compare other results normalized against it.Black lines in each panel show the prediction obtained from the simulations on Gaussian cores ($AR=1$; dotted lines) and filaments ($AR=\infty$, dashed lines), respectively. 
These new observational results are in close agreement with those in our simulations.
The ALMA 12m alone image (blue triangles) clearly depart from the expected values roughly following the trends identified in our simplified filament test cases (Sect.~\ref{sec:fil}). 
On the other hand, only after data combination, the ALMA 12m + IRAM-30m Feather data (yellow dots) is able to  recover the actual properties of filaments of different sizes FWHM$_0$.

Despite being overall in agreement with our filamentary predictions (dashed grey lines), our ALMA 12m alone results (blue triangles) show a significant scatter of a factor of $\sim$2. This effect might be partially explained by the different aspect ratio (AR; Sect.~\ref{sec:el_cores})  of the structures identified in OMC-3 (see Paper III) which would move these points upwards in these diagrams closer to the distribution expected for cores (dotted grey lines). The expected AR variations appear to be partially responsible of the observed variations in terms of flux recovery (top panel) and FWHM (bottom). Additional discrepancies are nonetheless expected given the simplicity of our synthetic toy models compared to the complex emission features seen in real ISM observations (see Paper I).

We remark here that all simulated and observed datasets are carried out using the ALMA 12m array in its most compact configuration (C43-1) in Band 3 (3mm). Although presenting significant issues, we note that these low-frequency observations in compact configurations represent the most favourable scenario to recover the extended emission seen in complex star-forming regions such as OMC-3. More severe issues are expected to affect observations at higher frequencies and/or in more extended configurations \citep[e.g.,][]{2023DiazGonzalez}. Additional work is needed to quantify these effects using different ALMA baselines and bands. However our study is sufficiently generic to state that any observation of objects with extended emission will benefit from combination with an observation with a large SD.

\section{Summary and conclusions}\label{sec:summary}

Our aim in this work was to investigate and quantify the impact of the ALMA instrumental response when characterizing the physical properties and gas organization of the ISM.
With cores and filaments at different scales, observations of star-forming regions are strongly affected by the interferometric filtering effect and especially by the short-spacing problem (Sect.~\ref{sec:intro}).
Although the scientific community is aware of these interferometric issues, the analysis of the results in the ALMA Science Archive suggest that data combination is still not fully implemented in recent ISM observations.
To explore the effects of the short-spacing information on the characterization of the physical structure of the ISM with ALMA, we investigated a series of CASA simulations (Sect.~\ref{sec:method}) and quantified the effects of data combination recovering the emission properties (total flux, peak flux, radial profile, and FWHM) of different core-like (Sect.~\ref{sec:cores}) and filamentary structures (Sect.~\ref{sec:fil}). We explored how the interferometric PSF (Sect.~\ref{sec:psf}) and the use of large SD telescopes (Sect.~\ref{sec:iram}) can affect these analysis. Finally, we also compared our synthetic observations with real ALMA data (Sect.~\ref{sec:real data}).

We summarize our main results as follows:
\begin{enumerate}
    \item We explore and compare targets showing different profiles (Gaussian and Plummer-like) and sizes defined by their FWHM, similar to observations.  
    Part of the target emission profile found at larger radii can easily exceed the maximum recoverable scale $\theta_{MRS}$ of the interferometer producing significant filtering effects. 
    As result, interferometric observations alone are not able to reproduce the properties of cores and filaments, even when their typical FWHM is below the MRS.
    
    \item The interferometric intrinsic filtering affects the quality of the observations in two ways: first, it produces large flux losses ($\sim$70-80$\%$ around the $\theta_{MRS}(12m)$ for Gaussian filaments) leading to strongly underestimated column densities and masses; and secondly, it systematically underestimates the FWHM of the object (by a factor of 2 around the $\theta_{MRS}(12m)$ for Gaussian filaments). As a result, sources observed with interferometers appear narrower and fainter than what they actually are. 

    \item The effect of the ALMA instrumental response depends on the geometry of the source. We demonstrated the filtering effects of the interferometer are much more severe on elongated objects with large aspect ratios $AR$ (filamentary) than in more symmetric Gaussian (core-like with most changes occurring in structures with $AR=2-3$.

    
    \item 
    The use of any technique of data combination allows us get closer to the real physical properties. 
    Among the data combination techniques explored in this work, the MACF procedure seems to give better results compared to feathering, although these differences are within $10\%$.
    Observers should use the full ALMA capabilities and apply for 12m+7m+TP observations.


    \item The combination of ALMA 12m array with a large SD (e.g. IRAM-30m) produce quantitatively better results than the ALMA (ACA) 7m + TP data, especially in the case of filamentary regions. The interferometric dataset will benefit from the combination even if the SD does not perfectly match the expected sensitivity of the ALMA 12m array observations.
    
\end{enumerate}

While explored in the case of ISM studies using compact ALMA configurations, these filtering effects are intrinsic to all interferometric observations. Additional studies are needed to quantify these effects in other science cases and ALMA configurations, however our study is sufficiently generic to state that any observation of objects with extended structures exceeding $\sim$0.5 the nominal MRS of the interferometer in size, will profit from combining with an observation with a large SD telescope. This work proves data combination as a necessary technique to recover the sky emission in ISM studies exploring complex star-forming regions, especially for nearby molecular clouds within 1~kpc distance such as Taurus and Orion. We will adopt these data techniques as standard procedure in future works of this EMERGE series (e.g., see Paper III).  

\begin{acknowledgements}
      This project has received funding from the European Research Council (ERC) under the European Union’s Horizon 2020 research and innovation programme (Grant agreement No. 851435).
      This paper makes use of the following ALMA data: ADS/JAO.ALMA\#2019.1.00641.S. ALMA is a partnership of ESO (representing its member states), NSF (USA) and NINS (Japan), together with NRC (Canada), MOST and ASIAA (Taiwan), and KASI (Republic of Korea), in cooperation with the Republic of Chile. The Joint ALMA Observatory is operated by ESO, AUI/NRAO and NAOJ.
      This work is based on IRAM-30m telescope observations carried out under project numbers 032-13, 120-20. IRAM is supported by INSU/CNRS (France), MPG (Germany), and IGN (Spain).
\end{acknowledgements}
\bibliographystyle{aa} 
\bibliography{paper} 

\appendix

\section{Observing the core mass function}\label{sec:cmf}

\begin{figure}[htbp]
	\centering
	\includegraphics[clip=true,trim=0.0cm 0.0cm 0.0cm 0.0cm,width=\columnwidth]{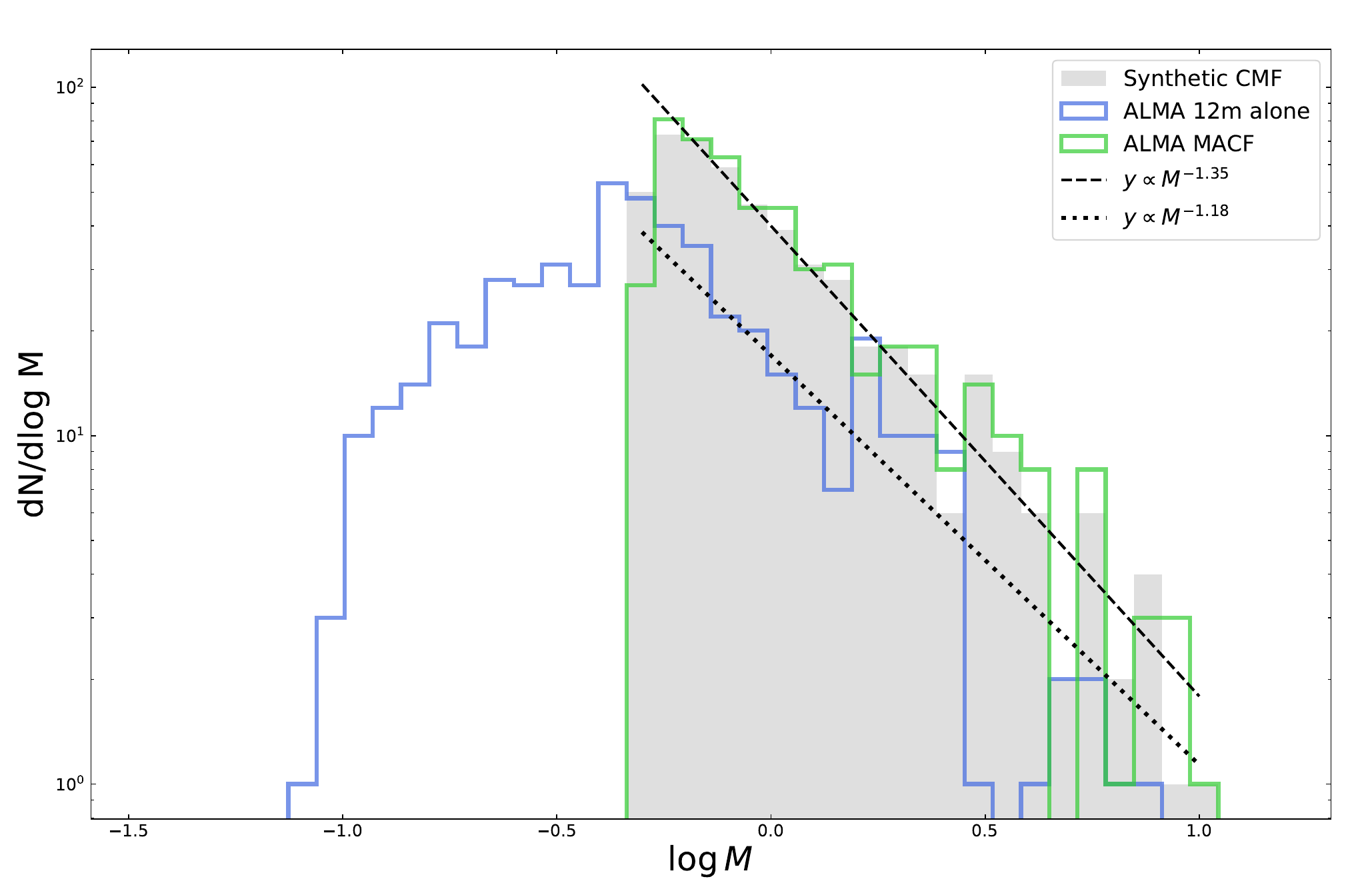}
	\caption{Analysis of the Core Mass Function (CMF). The synthetic Salpeter-like distribution used as input is marked by the grey-shadowed histogram.
    Blue and green line histograms describe the CMF recovered using the ALMA 12m alone and the ALMA MACF, respectively. The black dashed line displays the expected Salpeter dependence ($\alpha=1.35$), while the dotted line is the one that better describes the much shallower, top-heavy slope ($\alpha=1.18$) observed by the interferometer alone.}
	\label{fig:cmf}
\end{figure}

\begin{figure}[htbp]
	\centering
	\includegraphics[clip=true,trim=0.0cm 0.0cm 0.0cm 0.0cm,width=\columnwidth]{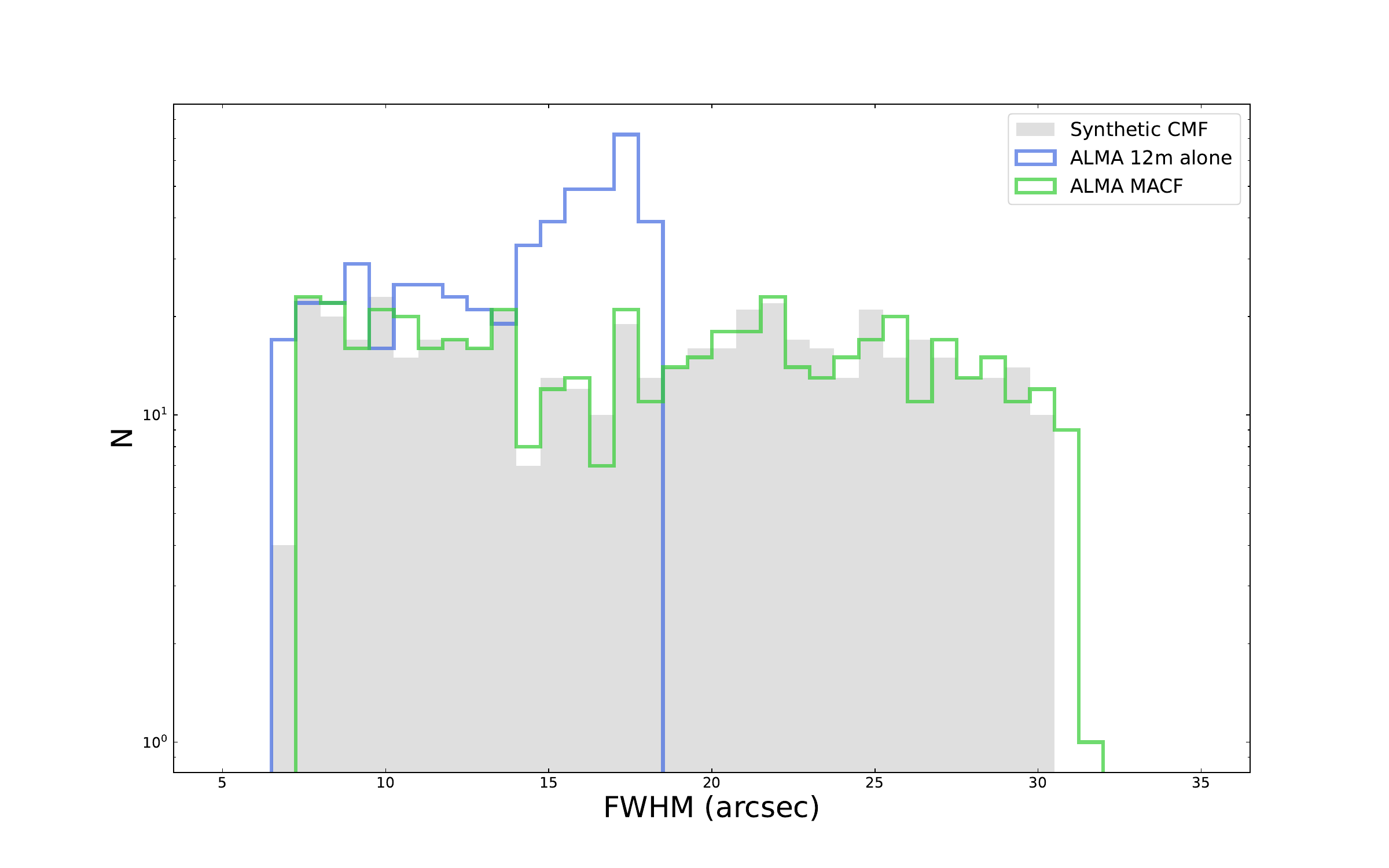}
	\caption{Analysis of the FWHM distribution recovered in the CMF.
    Blue and green line histograms describe the FWHM recovered using the ALMA 12m alone and the ALMA MACF, respectively.}
	\label{fig:cmf_fwhm}
\end{figure}

The Core Mass Function (CMF) describes the mass distributions of cores in molecular clouds prior the formation of stars. The CMF is a crucial diagnostic to investigate the origin of the Initial Mass Function \citep[IMF; see][]{1955Salpeter} and to test different theoretical models for star formation \citep[e.g.,][]{2001Kroupa,2003Chabrier}. 
The mass dependence of the CMF can be described by a power-law dependence such as
\begin{equation}
    \frac{dN}{dlog\ M}\propto M^{-\alpha}.
    \label{eq:cmf}
\end{equation}
Observational low-resolution studies in nearby low-mass, star-forming regions in the last two decades have revealed strong similarities between the slopes of CMF and IMF distributions \citep[both with $\alpha=1.35$; ][]{1955Salpeter} where the former is typically shifted a factor of 3 with respect to the latter, suggesting the IMF shape may be inherited from the CMF \citep[e.g.,][]{1998Motte,1998Testi,2007Alves,2015Konyves}. On the other hand, high-resolution ALMA observations of high-mass, star-forming regions report significantly shallower CMF slopes $\alpha<1$, known as top-heavy \citep{2018Motteb, 2019Kong}, challenging the direct relation between CMF and IMF.

Most of these recent ALMA studies are based on interferometric-only observations. 
Among others, the largest core survey to date has been obtained by the ALMA-IMF Large Program \citep{2022Motte}, a survey of 15 nearby massive protoclusters up to a resolution of $\sim0.01$~pc mapped in dust continuum at 1.3 and 3~mm. Approximately 700 cores have been detected in this ALMA-IMF program presented in \citet{2022Motte} and identified as Gaussian sources in their continuum maps \citep[see also][]{2023Pouteau}. ALMA-IMF only uses ALMA 12m alone data (ALMA 7m data exist, SD data not available for continuum) while combination was no possible due to issues of inconsistent quality across the sample \citep[see][]{2022Ginsburg}). 
Given the filtering effects seen in those Gaussian sources explored in Sect.~\ref{sec:cores}, it is crucial to assess the reliability of the CMF observed in these (and similar) ALMA studies. 


We aim to characterize the ALMA instrumental response on a standard CMF observed using the most compact configuration (C43-1) of the ALMA main array at a distance of Orion (D=414~pc). Our choices for both distance and telescope configuration maximize the interferometric recovery and are therefore meant to show the most favourable case for this type of ALMA studies \citep[e.g.,][]{2020Dutta}. We created a synthetic CMF drawing 500 core mass values following Eq.~\ref{eq:cmf} with a standard Salpeter-like slope of $\alpha=1.35$ within a mass range between $[0.5,10]\ M_\odot$, typical for dense cores, and assign them random FWHM values from a uniform distribution between 5 (similar to $\theta_{beam}(12m)$) and 30~arcsec ($\sim\theta_{MRS}(12m)$) which correspond to physical sizes of 0.01 and 0.06~pc at the selected distance. Given the mass M and the FWHM (in terms of spatial dispersion $\sigma=\frac{FWHM}{2\sqrt{2\ln{2}}}$) values for each core, we determined the associated H$_2$ column density (peak intensity of our cores) following
\begin{equation}
    N(H_2)=\frac{M}{m_H\ \mu_{H_2} \  2 \pi \sigma^2}
    \label{eq:columndensity}
\end{equation}
and convert them into their corresponding flux densities $F_\nu$ following Eq.~\ref{eq:flux}
assuming a homogeneous dust temperature of T=10K.

For each individual core we simulate their correspondingALMA 12m, 7m and TP observations following the same procedure already presented in Sect.~\ref{sec:method}. We then extracted the individual peak intensity, FWHM, and masses of each target 
and investigate the resulting combined CMF after applying different data combination techniques.

We display the distinct recovered CMFs (line histograms) 
and compare them with our input distribution (grey histogram) in Fig.~\ref{fig:cmf}.
The black dotted and dashed lines included in the plot do not represent a fit of the distributions, but only representative slopes that can describe the data to better compare them. The CMF observed using only the ALMA 12m alone (blue histogram) shows a shallower distribution ($\alpha=-1.18$) shifted towards lower masses with respect to the input one (grey shaded area with $\alpha=-1.35$). 
This result can be explained as a combination of the flux losses and the FWHM deviations reported for individual Gaussians in Sect.~\ref{sec:cores}, leading to the systematic underestimation of all core masses in large statistical samples producing apparent top-heavy CMF. 

As primary driver of the above deviations, we illustrate the distribution of FWHM in Fig.~\ref{fig:cmf_fwhm} (changes in flux peak are minor and thus not shown here). The distribution recovered by the ALMA 12m alone (blue histogram) is truncated at $\sim18$~arcsec with respect to the original (grey histogram) and the one recovered by the ALMA MACF method (green histogram). In agreement with Fig.~\ref{fig:cores_properties} (lower panel), the recovered FWHM of cores $\gtrsim20$~arcsec is systematically narrower than expected.
Moreover, these interferometric effects could also effectively reduce the detection rates in ALMA core surveys since many targets could be shifted below the mass completeness threshold of these observations \citep[e.g., $\sim 1$~M$_\odot$ in][]{2023Pouteau}.
Despite their simplicity, our mock observations demonstrate how observational artefacts such as interferometric filtering can critically bias the estimates of the CMF slope even in the most compact ALMA  configurations.

Our findings reinforce the conclusions drawn by \citet{2023Padoan} using more realistic models of clouds and radiative transfer calculations. These authors found that core masses inferred from ALMA observations, and thus the observed CMF, are highly unreliable due to the combination of projection, filtering, and temperature effects. As result, \citet{2023Padoan} recover a shallower power-law tail but shifted instead towards higher masses which the authors attribute to the additional effects of background subtraction. Our simulations demonstrate how most of these observational biases are already present under the idealized conditions and can be directly attributed to pure interferometric filtering. 

Compared to the above interferometric-alone results, we display the CMF recovered by our ALMA MACF observations in Fig.~\ref{fig:cmf} (green distribution). The simulations illustrates the improvement on the overall flux (and therefore mass) recovery in our sample. The resulting CMF obtained using the ALMA MACF combination reproduces (within the noise) the expected Salpeter-like slope across the entire dynamic range of masses of interest. Our results demonstrate how the combination of zero-spacing information is crucial for obtaining reliable estimates of the core masses in star-forming regions observed with interferometers even when these cores are partially unresolved.

\clearpage

\twocolumn[\section{Elongated geometries: synthetic observations}]

\begin{figure*}[b]
\centering
	\includegraphics[clip=true,trim=0.0cm 0.0cm 0.0cm 0.0cm,width=0.82\textwidth]{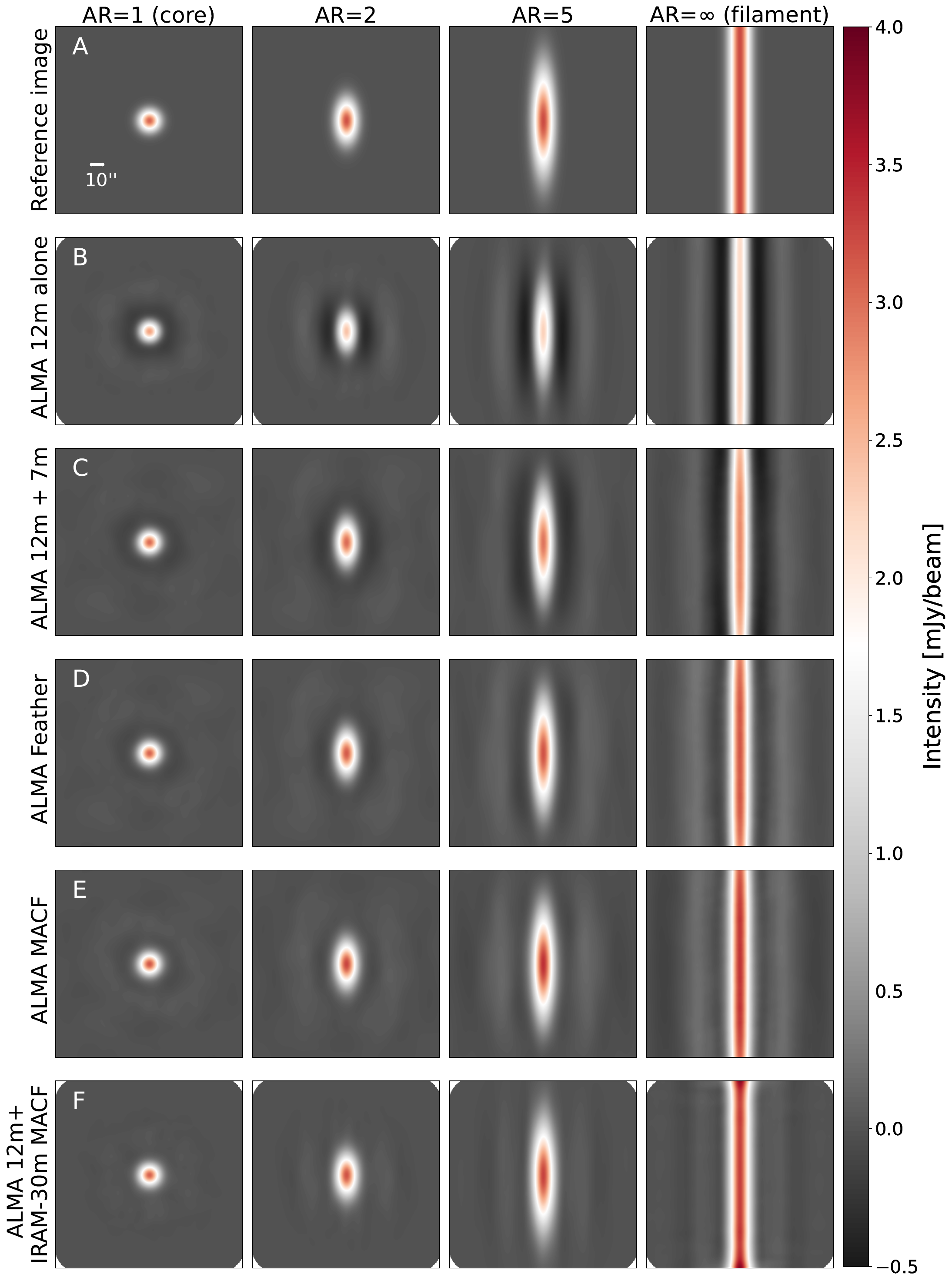}
	\caption{Noise-free simulations of isolated objects with FWHM$_0$=15~arcsec and different aspect ratios AR. From left to right: AR=1 (core), 2, 5, and $\infty$ (infinite filament). From top to bottom: reference model (panel A), and simulated ALMA 12m alone (panel B), ALMA 12m + 7m (panel C), ALMA Feather (panel D), ALMA MACF (panel E), and ALMA 12m + IRAM-30m MACF (panel F) observations, respectively. }
	\label{fig:elcores_comparison}
\end{figure*}

\end{document}